# Unifying Brillouin scattering and cavity optomechanics

Raphaël Van Laer,* Roel Baets and Dries Van Thourhout
*Photonics Research Group, Ghent University–imec, Belgium*
*Center for Nano- and Biophotonics, Ghent University, Belgium*

So far, Brillouin scattering and cavity optomechanics were mostly disconnected branches of research – although both deal with photon-phonon coupling. This begs for the development of a broader theory that contains both fields. Here, we derive the dynamics of optomechanical cavities from that of Brillouin-active waveguides. This explicit transition elucidates the link between phenomena such as Brillouin amplification and electromagnetically induced transparency. It proves that effects familiar from cavity optomechanics all have traveling-wave partners, but not vice versa. We reveal a close connection between two parameters of central importance in these fields: the Brillouin gain coefficient and the zero-point optomechanical coupling rate. This enables comparisons between systems as diverse as ultracold atom clouds, plasmonic Raman cavities and nanoscale silicon waveguides. In addition, back-of-the-envelope calculations show that unobserved effects, such as photon-assisted amplification of traveling phonons, are now accessible in existing systems. Finally, we formulate both circuit- and cavity-oriented optomechanics in terms of vacuum coupling rates, cooperativities and gain coefficients, thus reflecting the similarities in the underlying physics.

## I. INTRODUCTION

Brillouin scattering [1] and cavity optomechanics [2] have been intensively studied in recent years. Both concern the interaction between light and sound, but they were part of separate traditions. Already in the early 1920s, diffraction of light by sound was studied by Léon Brillouin. Therefore, such inelastic scattering is called *Brillouin scattering* [3, 4]. The effect is known as *stimulated* Brillouin scattering (SBS) [5–7] when a strong intensity-modulated light field generates the sound, often with classical applications such as spectral purification [8] and microwave signal processing [9] in mind. In contrast, cavity optomechanics arose from Braginsky's efforts to understand the limits of gravitational wave detectors in the 1970s – and greatly expanded since the demonstration of phonon lasing in microtoroids [10]. By and large, it aims to control both optical and mechanical quantum states [11–13].

Historically, a number of important differences hindered their merger. For instance, SBS generally dealt with high-group-velocity and cavity optomechanics with low-group-velocity acoustic phonons. In addition, bulk electrostrictive forces usually dominated phonon generation in SBS – while radiation pressure at the boundaries took this role in cavity optomechanics. Further, cavity optomechanics typically studied resonators with much lower phonon than photon dissipation – whereas Brillouin lasers [8, 14, 15] operate in the reversed regime [16]. Finally, SBS is often studied not in cavities but in optically broadband waveguides [1]. Thus, particular physical systems used to be firmly placed in either one or the other research paradigm.

Lately, the idea that these are mostly superficial classifications has been gaining traction. Indeed, in both cases light generates motion and the motion phase-modulates light. Next, this spatiotemporal phase-modulation creates motional sidebands – which interfere with those initially present. The research fields share this essential feedback loop. Some connections have already been made. For instance, electrostrictive forces were exploited for sideband cooling [17, 18] and induced transparency [19, 20] while radiation pressure contributed to SBS in dual-web fibers [21] and silicon waveguides [22–25].

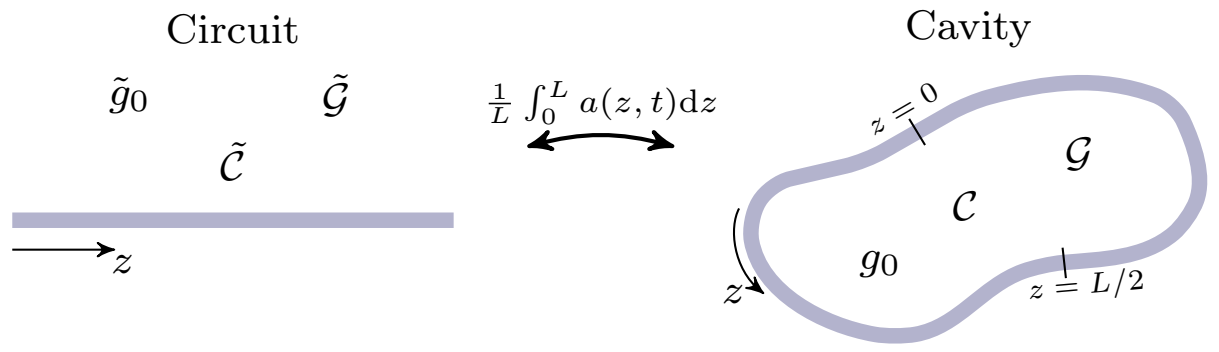

FIG. 1. **From circuit to cavity optomechanics.** We explicitly derive the physics of optomechanical cavities (right) from that of Brillouin-active waveguides (left). Therefore, both traveling-wave and cavity-based optomechanics can be cast in terms of vacuum coupling rates ($\tilde{g}_0$ and $g_0$), cooperativities ($\tilde{\mathcal{C}}$ and $\mathcal{C}$) and gain coefficients ($\tilde{\mathcal{G}}$ and $\mathcal{G}$).

In this work, we derive the dynamics of optomechanical cavities from that of Brillouin-active waveguides (fig.1). The transition holds for both co- and counter-propagating pump and Stokes waves, i.e. for "forward" and "backward" scattering, and for opto-acoustic coupling between two different or two identical optical modal fields, i.e. for "inter-" [18, 26–29] or "intra-" [22, 23, 28, 30–35] modal scattering (fig.2). Hence, all flavours of photon-phonon interaction are treated on the same footing. Moreover, this spatially averaged cavity dynamics is found to be equivalent to the standard Hamiltonian of cavity optomechanics [2] – even in the case of low-finesse phonons. It turns out that this cavity dynamics can be mapped – by swapping space and time ($z \leftrightarrow t$) – on the steady-state spatial evolution of the opto-acoustic fields in the waveguide. The treatment describes Raman effects as well.

* raphael.vanlaer@intec.ugent.be

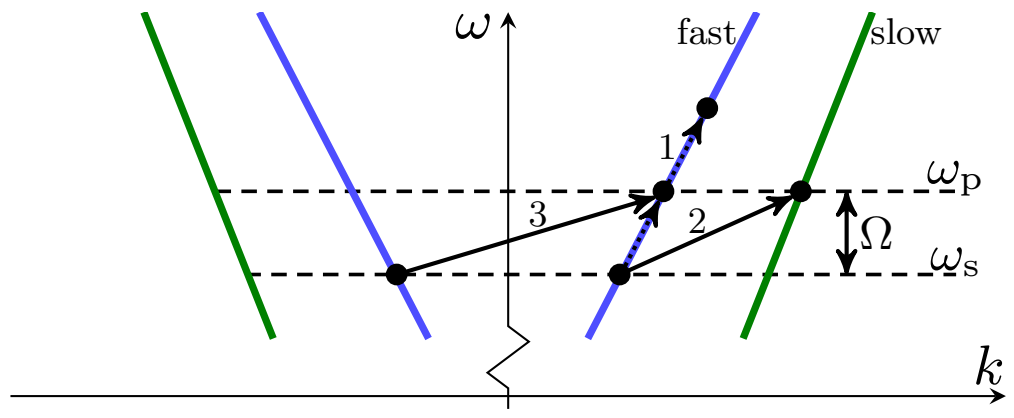

FIG. 2. **Phase-matching diagrams.** The optical dispersion relation $\omega(k)$ shows phonon-mediated coupling between co- or counter-propagating photons and between two identical (intra-) or two different (inter-modal) optical modes. Therefore, there are generally four types of optomechanics, of which three are indicated in this diagram. The fourth is counter-coupling between two different optical modes.

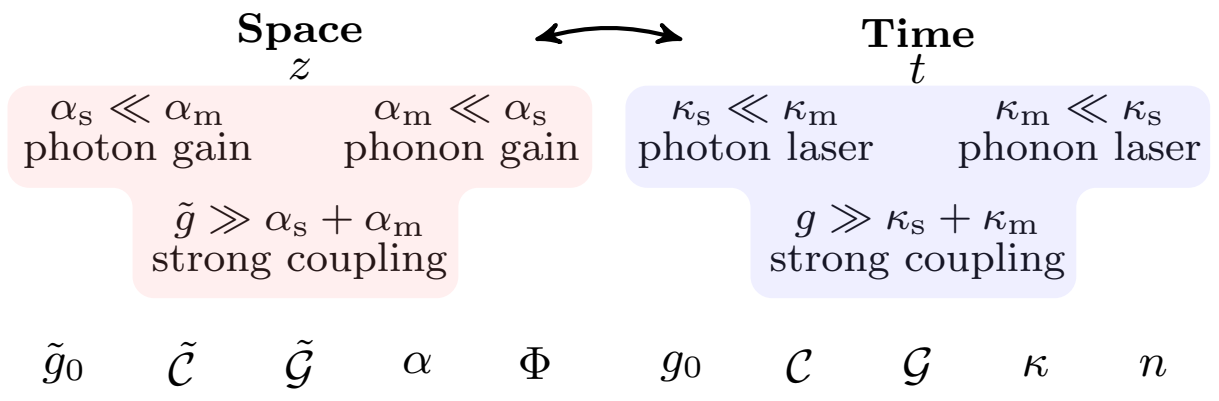

FIG. 3. **Symmetry of circuit and cavity optomechanics.** Each temporal optomechanical effect has a spatial symmetry partner. Thus, the description of these effects can be cast in terms of conceptually similar figures of merit. The scheme assumes a red-detuned optical probe; "gain" and "laser" should respectively be replaced by "loss" and "cooling" for a blue-detuned optical probe. The meaning of the figures of merit is discussed in the main text.

This implies that the plethora of optomechanical effects, such as stimulated Brillouin scattering [3, 4, 36], slow light [37–39], optomechanically induced transparency [39, 40], ground-state cooling [11, 41] etc., are different aspects of the same feedback loop. The rigorous transition decisively indicates that both fields are a subset of a larger theory of photon-phonon interaction, which may be built on a single Hamiltonian [42]. This is not to say that they are identical: a Brillouin-active waveguide supports complex spatiotemporal phenomena [43–45] and noise dynamics [46, 47] not present in a high-finesse optomechanical cavity. Nevertheless, in the resulting picture (fig.3), both traveling-wave and cavity-based photon-phonon interaction can be classified according to (1) the damping hierarchy of the photons and phonons and (2) the strength of the photon-phonon coupling with respect to the largest dissipation channel. For weak coupling, the long-lived particle species – either photons or phonons – triggers the photon-phonon conversion. The short-lived particle species cannot truly build up and is thus slaved to its long-lived partner; it is merely created in short segments (of space or time) and immediately decays afterwards.

All Brillouin-active waveguides so far exhibited far stronger phononic than photonic propagation losses; in addition, the coupling was always weak relative to this phononic damping. Hence, there are two to date unexplored regimes of guided-wave optomechanics: (1) photon-assisted amplification of traveling phonons and (2) strong coupling between traveling photons and phonons (fig.3). The strong coupling regime produces either traveling entangled photon-phonon pairs or state swapping between light and sound along the waveguide, depending on the details (e.g. probe detuning) of the experiment. Although currently unobserved, both effects may be an asset in future quantum phononic networks [13, 48–51]. For instance, in the strong coupling regime the flying phonon – entangled to its photonic partner – could be detected piezo-electrically or optically and thereby enable Bell tests [52–54] between two different particle species. Our back-of-the-envelope estimates show that these regimes can be achieved in existing systems, such as dual-web fibers and silicon nanowaveguides.

The transition (fig.1) assumes that the photonic and phononic modes of the waveguide are not disturbed too strongly by looping it into a cavity. This is justified in many cases since cavity designs aim to minimize the losses (e.g. due to bending) induced by any modal perturbations. Within this approximation, it permits translations between circuit- and cavity-oriented figures of merit. For instance, we identify a connection between the Brillouin or Raman gain coefficient $\tilde{\mathcal{G}}$ and the zero-point coupling rate $g_0$. The former ($\tilde{\mathcal{G}}$) quantifies the pump power and waveguide length required to amplify a Stokes seed appreciably [3, 4]. The latter ($g_0$) captures the interaction strength between a single photon and a single phonon in an optomechanical cavity [2]. The transition proves that these figures of merit are inextricably linked by

$$v_{\rm p} v_{\rm s} \frac{(\hbar\omega_{\rm p})\,\Omega_{\rm m}}{4L} \left(\frac{\tilde{\mathcal{G}}}{Q_{\rm m}}\right) = g_0^2 \tag{1}$$

with $v_{\rm p}$ and $v_{\rm s}$ the group velocities of the pump and Stokes waves, $\hbar\omega_{\rm p}$ the pump photon energy, $\frac{\Omega_{\rm m}}{2\pi}$ the mechanical resonance frequency, $L$ the cavity roundtrip length and $Q_{\rm m}$ the waveguide's mechanical quality factor. This link is independent of the type of driving optical force and of the relative photon and phonon damping. Similarly, we derive connections between each of the circuit- and cavity-oriented figures of merit: between the vacuum coupling rates ($\tilde{g}_0$ and $g_0$, see (21)), the cooperativities ($\tilde{\mathcal{C}}$ and $\mathcal{C}$, see (34)) and the gain coefficients ($\tilde{\mathcal{G}}$ and $\mathcal{G}$, see (36)).

Notably, this treatment goes beyond cavity optomechanical systems that have a clear circuit equivalent (as in fig.1). Indeed, the standard cavity Hamiltonian $\hat{\mathcal{H}} = \hbar\omega_{\rm c}(\hat{x})\hat{a}^\dagger\hat{a} + \hbar\Omega_{\rm m}\hat{b}^\dagger\hat{b}$ [2] also captures the temporal dynamics of cavity optomechanics based on Bose-Einstein condensates [55, 56] or plasmonic Raman cavities [57]. The physics of all these diverse systems can be understood in the scheme of fig.3. On top of the similar dynamics, this means that the photon-phonon interaction efficiency of a larger class of systems can now be compared in a single framework. For instance, the gain coefficient of a silicon nanowire can be converted to the vacuum coupling rate of a hypothetical cavity (through (1)); which can next be compared to that of any other cavity optomechanical system. In reverse, the link enables the conversion of a vacuum coupling rate of an actual cavity optomechanical system into a hypothetical guided-wave coupling rate (through (21)); which can next be compared to that of any other waveguide. We give examples of such conversions, which can be tested empirically in many cases, in section V.

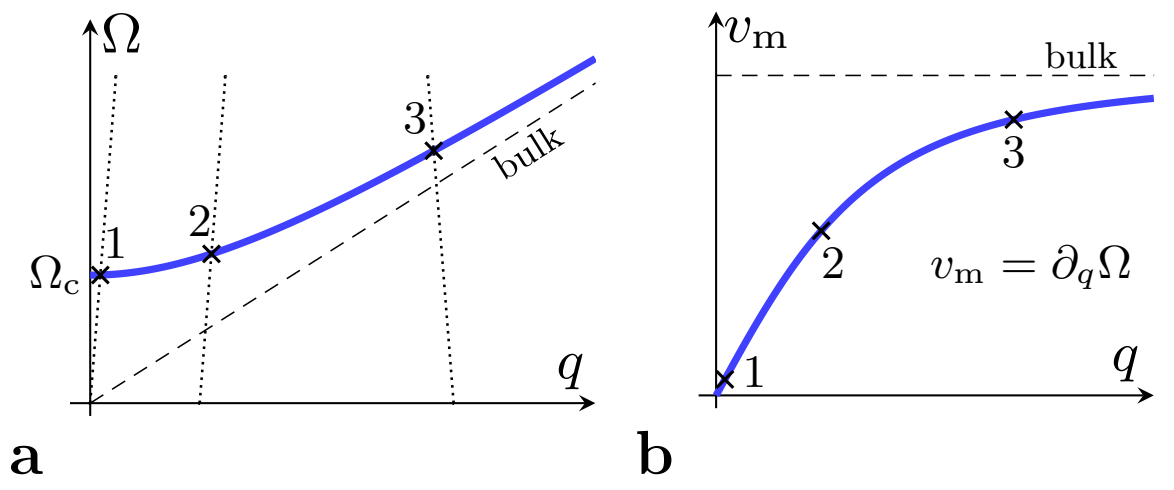

FIG. 4. **Example of phonon dispersion relation.** **a**, The frequency $\Omega(q)$ of transversally trapped acoustic phonons has a Raman-like cut-off $\Omega_{\rm c}$ for low $q$ and approaches the bulk relation for large $q$. **b**, Thus, the phonon group velocity $v_{\rm m}$ vanishes for low $q$ and becomes the bulk speed for large $q$.

The paper is organized as follows: in section II we describe a minimal model of traveling-wave optomechanics and frame it in terms of a guided-wave vacuum coupling rate $\tilde{g}_0$ and cooperativity $\tilde{\mathcal{C}}$. Next, we make the mean-field transition to a cavity in section III. At that point, we also discuss the limitations of the analysis. The resulting dynamical effects are treated in section IV. The prospects for observing new effects are considered in section V and we conclude in section VI.

## II. CIRCUIT OPTOMECHANICS

In particular, we study the interaction between a pump field with envelope $a_{\rm p}(z,t)$ and a red-detuned Stokes field envelope $a_{\rm s}(z,t)$ mediated by an acoustic field with envelope $b(z,t)$. The envelopes contain only the slowly varying part of the photonic-phononic fields; rapidly oscillating factors $e^{i(kz-\omega t)}$ were removed in each case. The guided optical modes correspond to the points $(\omega_{\rm p}, k_{\rm p})$ and $(\omega_{\rm s}, k_{\rm s})$ in the optical dispersion relation (fig.2). By energy and wave-momentum [58] conservation, the excited phonon has an angular frequency $\Omega = \omega_{\rm p} - \omega_{\rm s}$ and wavevector $q = k_{\rm p} \mp k_{\rm s}$. The nature of the optical modes (co/counter and fast/slow) and the acoustic dispersion relation determine the wavevector $q$ and group velocity $v_{\rm m}$ of the excited phonons (fig.2&4).

Traveling-wave photon-phonon coupling is governed by the following dynamical evolution [42, 47, 59]

$$\begin{aligned} v_{\rm p}^{-1}\partial_t a_{\rm p} + \partial_z a_{\rm p} &= -i\tilde{g}_0 a_{\rm s} b - \tilde{\chi}_{\rm p}^{-1} a_{\rm p} \\ v_{\rm s}^{-1}\partial_t a_{\rm s} \pm \partial_z a_{\rm s} &= -i\tilde{g}_0 b^\dagger a_{\rm p} - \tilde{\chi}_{\rm s}^{-1} a_{\rm s} \\ v_{\rm m}^{-1}\partial_t b + \partial_z b &= -i\tilde{g}_0 a_{\rm s}^\dagger a_{\rm p} - \tilde{\chi}_{\rm m}^{-1} b \end{aligned} \tag{2}$$

Its derivation proceeds from Maxwell's and the elasticity equations on the assumption that the envelopes vary slowly in space and time [42, 47, 59]. This starting point and the following treatment holds quantum mechanically if one takes care to treat the envelopes in (2) as operators [42, 47] obeying the equal-time commutator

$$\left[a_\gamma(z,t), a_{\gamma'}^\dagger(z',t)\right] = \sqrt{v_\gamma v_{\gamma'}}\,\delta_{\gamma\gamma'}\delta(z-z') \tag{3}$$

with "$\gamma$" an index running over the pump "p", Stokes "s" and mechanical wave "m", $v_\gamma$ the group velocities, $\delta_{\gamma\gamma'}$ the Kronecker delta, $\delta(z)$ the Dirac delta distribution and $a_{\rm m} = b$ for notational convenience. We flux-normalized the field operators $a_\gamma$ such that $\Phi_{\rm p} = a_{\rm p}^\dagger a_{\rm p}$, $\Phi_{\rm s} = a_{\rm s}^\dagger a_{\rm s}$ and $\Phi_{\rm m} = b^\dagger b$ correspond to the number of pump photons, Stokes photons and phonons passing through a cross-section of the waveguide per second. We will treat highly occupied (i.e. large mean flux $\langle\Phi_\gamma\rangle$) modes as classical in the remainder of the paper, as is standard [2, 60–62]. Further, we denote $\tilde{g}_0$ the traveling-wave vacuum coupling rate (to be discussed further on), $\tilde{\chi}_\gamma^{-1} = \frac{\alpha_\gamma}{2} - i\tilde{\Delta}_\gamma$ the susceptibilities, $\alpha_\gamma$ the propagation losses and $\tilde{\Delta}_\gamma$ the wavevector offsets between externally applied fields and the intrinsic waveguide modes.

In some systems, e.g. for the Raman-like low-group-velocity phonons (fig.4) associated with forward intra-modal scattering [22–24, 28, 30], the phonon wavelength $\frac{2\pi}{q}$ can be substantially larger than its decay length $\alpha_{\rm m}^{-1}$ – so its slowly-varying amplitude treatment breaks down. Then the acoustic excitation is better treated as a localized series of mechanical oscillators [23, 30, 59], essentially dealing with each cross-sectional slice of the waveguide as an artificial Raman-active molecule. The above dynamics (2), however, contains these systems as well by letting the phonon decay length $\alpha_{\rm m}^{-1}$ vanish. Further, the sign ($\pm$) in the Stokes equation indicates the difference between forward (+) and backward (−) photon-phonon coupling. Cascaded scattering [30, 63] and noise [46, 47] can and should be added to this model in some instances. In fact, (2) can be regarded as the unique, minimal model for guided-wave Brillouin scattering [42, 47, 59]. We discuss potential extensions in section V; in the following, we only need the minimal model (2), future extended versions can be dealt with similarly.

The Manley-Rowe relations [3, 64] guarantee that a single, unique figure of merit $\tilde{g}_0$ captures all conservative optical forces and scattering. Indeed, in the lossless case ($\alpha_\gamma = 0$), the rate of pump photon destruction must equal the rate of Stokes photon and phonon creation:

$$-\partial_z\Phi_{\rm p} = \pm\partial_z\Phi_{\rm s} = \partial_z\Phi_{\rm m} = -\tilde{g}_0\left(ia_{\rm s}^\dagger b^\dagger a_{\rm p} + \text{h.c.}\right) \tag{4}$$

Similar to $g_0$ in a cavity [2], $\tilde{g}_0$ quantifies the interaction strength between a single photon and a single phonon, but in this case flying along a waveguide instead of trapped in a cavity. We take $\tilde{g}_0$ real and positive without loss of generality. Briefly specializing to forward intra-modal scattering, the mean-field transition of section III will show that (see Appendices)

$$\tilde{g}_0 = -\tilde{x}_{\mathrm{ZPF}} \left.\frac{\partial k_{\mathrm{p}}}{\partial x}\right|_{\omega_{\mathrm{p}}} \tag{5}$$

with

$$\tilde{x}_{\mathrm{ZPF}} = x_{\mathrm{ZPF}}\sqrt{\frac{\delta L}{v_{\mathrm{m}}}} = \sqrt{\frac{\hbar}{2m_{\mathrm{eff}}v_{\mathrm{m}}\Omega_{\mathrm{m}}}} \tag{6}$$

the guided-wave zero-point motion and $m_{\mathrm{eff}}$ the effective mass of the mechanical mode per unit length. Indeed, a short waveguide section of length $\delta L$ contains $\langle n_{\mathrm{m}}\rangle = \frac{\delta L}{v_{\mathrm{m}}}\langle \Phi_{\mathrm{m}}\rangle$ phonons with $\langle \Phi_{\mathrm{m}}\rangle$ the mean phonon flux. As particle fluxes – instead of numbers – are fundamental in the traveling-wave Manley-Rowe relations (4), the zero-point motion is rescaled by precisely this factor $(\delta L/v_{\mathrm{m}})^{1/2}$ relative to the actual zero-point motion $x_{\mathrm{ZPF}}$ [2] of the $\delta L$-section

$$x_{\mathrm{ZPF}} = \sqrt{\frac{\hbar}{2m_{\mathrm{eff}}\delta L\Omega_{\mathrm{m}}}} \tag{7}$$

Therefore, the traveling-wave vacuum coupling rate $\tilde{g}_0$ is determined by the wavevector shift induced by mechanical motion at fixed frequency, while the cavity vacuum coupling rate $g_0$ is determined by the frequency shift induced by mechanical motion at fixed wavevector [2]. Notably, the interpretation of $\tilde{g}_0$ as the coupling strength between a single traveling photon and phonon holds also for inter-modal and backward scattering (see Appendices).

In steady-state ($\partial_t \to 0$) and for a constant, strong pump ($\Phi_{\mathrm{p}}(z) = \Phi_{\mathrm{p}}(0)$), the evolution (2) reduces to

$$\begin{aligned} \partial_z a_{\mathrm{s}} &= \mp i\tilde{g}_0 b^\dagger a_{\mathrm{p}} \mp \tilde{\chi}_{\mathrm{s}}^{-1} a_{\mathrm{s}} \\ \partial_z b &= -i\tilde{g}_0 a_{\mathrm{s}}^\dagger a_{\mathrm{p}} - \tilde{\chi}_{\mathrm{m}}^{-1} b \end{aligned} \tag{8}$$

The phonon decay length $\alpha_{\mathrm{m}}^{-1}$ is generally largest for backward scattering. Even then, it typically does not exceed $\alpha_{\mathrm{m}}^{-1} \sim 100\,\mu\mathrm{m}$ [3, 65]. Therefore, the photon decay length massively exceeds the phonon decay length in Brillouin-active waveguides to date ($\alpha_{\mathrm{s}} \ll \alpha_{\mathrm{m}}$). A full solution of (8) exists but yields little intuitive insight (see Appendices). Therefore, we initially focus on two subcases: the conventional ($\alpha_{\mathrm{s}} \ll \alpha_{\mathrm{m}}$) and the reversed case ($\alpha_{\mathrm{m}} \ll \alpha_{\mathrm{s}}$), both in the weak coupling regime ($\tilde{g}_0\sqrt{\Phi_{\mathrm{p}}} \ll \alpha_{\mathrm{s}} + \alpha_{\mathrm{m}}$). These examples illustrate how one can formulate guided-wave optomechanics, including the classical stimulated Brillouin regime, in terms of the vacuum coupling rate $\tilde{g}_0$ and cooperativity $\tilde{\mathcal{C}}$.

First, strongly damped phonons ($\alpha_{\mathrm{s}} \ll \alpha_{\mathrm{m}}$) act as a localized slave wave ($\partial_z b \to 0$) given by $b = -i\tilde{\chi}_{\mathrm{m}}\tilde{g}_0 a_{\mathrm{s}}^\dagger a_{\mathrm{p}}$. On resonance ($\tilde{\Delta}_\gamma = 0$), we thus have

$$\partial_z a_{\mathrm{s}} = \mp(1-\tilde{\mathcal{C}})\frac{\alpha_{\mathrm{s}}}{2} a_{\mathrm{s}} \tag{9}$$

with

$$\tilde{\mathcal{C}} = \frac{4\tilde{g}_0^2\Phi_{\mathrm{p}}}{\alpha_{\mathrm{s}}\alpha_{\mathrm{m}}} = \frac{4\tilde{g}^2}{\alpha_{\mathrm{s}}\alpha_{\mathrm{m}}} \tag{10}$$

the guided-wave cooperativity and $\tilde{g} = \tilde{g}_0\sqrt{\Phi_{\mathrm{p}}}$ the pump-enhanced spatial coupling rate. Therefore, $\tilde{\mathcal{C}} = 1$ is the threshold for net phonon-assisted gain on flying photons. Since $P_{\mathrm{p}} = \hbar\omega_{\mathrm{p}}\Phi_{\mathrm{p}}$ is the pump power, we obtain $\tilde{\mathcal{C}} = \frac{\tilde{\mathcal{G}}P_{\mathrm{p}}}{\alpha_{\mathrm{s}}}$ and

$$\tilde{\mathcal{G}} = \frac{4\tilde{g}_0^2}{\hbar\omega_{\mathrm{p}}\alpha_{\mathrm{m}}} \tag{11}$$

the well-known Brillouin gain coefficient [3, 4], here framed in terms of a spatial coupling rate $\tilde{g}_0$ and cooperativity $\tilde{\mathcal{C}}$. It characterizes the spatial exponential build-up of a Stokes seed in case of highly damped phonons ($\alpha_{\mathrm{s}} \ll \alpha_{\mathrm{m}}$). Since $\langle \Phi_{\mathrm{m}}\rangle = \frac{\alpha_{\mathrm{s}}}{\alpha_{\mathrm{m}}}\tilde{\mathcal{C}}\langle \Phi_{\mathrm{s}}\rangle \ll \langle \Phi_{\mathrm{s}}\rangle$, there are on average far fewer phonons than photons flying along the waveguide in this case. The system enters the strong coupling regime as soon as $\tilde{\mathcal{C}} \sim \frac{\alpha_{\mathrm{m}}}{\alpha_{\mathrm{s}}}$ (see section IV).

Second, when the phononic damping is lowest ($\alpha_{\mathrm{m}} \ll \alpha_{\mathrm{s}}$), we similarly get a slaved Stokes wave ($\partial_z a_{\mathrm{s}} \to 0$) given by $a_{\mathrm{s}} = -i\tilde{\chi}_{\mathrm{s}}\tilde{g}_0 b^\dagger a_{\mathrm{p}}$ resulting in ($\tilde{\Delta}_\gamma = 0$)

$$\partial_z b = -(1-\tilde{\mathcal{C}})\frac{\alpha_{\mathrm{m}}}{2} b \tag{12}$$

such that $\tilde{\mathcal{C}} = 1$ also yields the threshold for net photon-assisted gain on flying phonons. Since $\langle \Phi_{\mathrm{s}}\rangle = \frac{\alpha_{\mathrm{m}}}{\alpha_{\mathrm{s}}}\tilde{\mathcal{C}}\langle \Phi_{\mathrm{m}}\rangle \ll \langle \Phi_{\mathrm{m}}\rangle$, there are far fewer photons than phonons flying along the waveguide in this case. The system enters the strong coupling regime as soon as $\tilde{\mathcal{C}} \sim \frac{\alpha_{\mathrm{s}}}{\alpha_{\mathrm{m}}}$. By replacing the undepleted pump with an undepleted, strong Stokes mode ($\tilde{g} = \tilde{g}_0\sqrt{\Phi_{\mathrm{s}}}$), it follows that an anti-Stokes seed sees larger loss by a factor $(1+\tilde{\mathcal{C}})$ conventionally ($\alpha_{\mathrm{s}} \ll \alpha_{\mathrm{m}}$) and that a guided-wave phonon channel can be cooled by a factor $(1+\tilde{\mathcal{C}})$ when it has the lowest propagation loss ($\alpha_{\mathrm{m}} \ll \alpha_{\mathrm{s}}$). An undepleted, strong phononic beam ($\tilde{g} = \tilde{g}_0\sqrt{\Phi_{\mathrm{m}}}$) yields similar coupling between the pump and Stokes wave.

The coupling rate $\tilde{g}$ and the cooperativity $\tilde{\mathcal{C}}$ respect the symmetry between flying photons and phonons, whereas the gain coefficient $\tilde{\mathcal{G}}$ (11) is most relevant in case of stronger phonon damping. Therefore, we regard $\tilde{g}$ and $\tilde{\mathcal{C}}$ as more natural and fundamental figures of merit. It is straightforward to extend the above discussion for absorptive decay of the pump flux, i.e. $\Phi_{\mathrm{p}}(z) = \Phi_{\mathrm{p}}(0)e^{-\alpha_{\mathrm{p}}z}$ and non-zero wavevector detunings $\tilde{\Delta}_\gamma \neq 0$.

So far, we discussed two subcases of guided-wave Brillouin scattering. We treat the strong coupling regime in section IV and the full solution in the Appendices. Next, we move on to cavity optomechanics via the mean-field transition.

## III. BRIDGE TO CAVITY OPTOMECHANICS

In this section, we transition to an optical cavity – made from a Brillouin-active waveguide – of roundtrip length $L$ (fig.1). To do so, we introduce the *mean-field* envelope operators

$$\overline{a}(t) = \frac{1}{L}\int_0^L a(z,t)\mathrm{d}z \tag{13}$$

for both the optical ($\overline{a}_{\mathrm{p/s}}(t)$) and acoustic ($\overline{b}(t)$) fields. Such mean-field models have found early use in the treatment of fluorescence [66] and recently also in the context of frequency combs [67]. During roundtrip propagation, each field obeys dynamics of the form (see (2))

$$v^{-1}\partial_t a + \partial_z a = \zeta - \tilde{\chi}^{-1} a \tag{14}$$

with $\zeta$ the nonlinear interaction term. To describe the cavity feedback (fig.1), we add the boundary condition

$$a(0,t) = \sqrt{1-\alpha'}\sqrt{1-\mu}\, e^{i\varphi} a(L,t) + \sqrt{\mu}\, s(t) \tag{15}$$

with $\alpha'$ the additional loss fraction along a roundtrip (on top of $\alpha$, such as bending losses), $\mu$ the fraction of photons or phonons coupled to an in- or output channel, $\varphi$ the roundtrip phase shift and $s(t)$ the flux-normalized envelope of injected photons or phonons. By Taylor-expansion of (15), we get

$$a(L,t) - a(0,t) \approx \left(\frac{\alpha' + \mu}{2} - i\varphi\right)\overline{a}(t) - \sqrt{\mu}\, s(t) \tag{16}$$

with higher-order terms negligible and $a(L,t) \approx \overline{a}(t)$ in the high-finesse limit. Low-finesse situations, particularly relevant for phonons, are treated further on (see (24)). We operate close to the cavity resonance, such that $\varphi \ll 2\pi$. Next, we let (13) operate on (14) and use $\overline{\partial_t a} = \dot{\overline{a}}(t)$:

$$v^{-1}\dot{\overline{a}}(t) + L^{-1}\{a(L,t) - a(0,t)\} = \overline{\zeta}(t) - \tilde{\chi}^{-1}\overline{a}(t) \tag{17}$$

We insert (16) in (17) and find

$$\dot{\overline{a}} = v\overline{\zeta} - \chi^{-1}\overline{a} + \frac{\sqrt{\mu}}{T}s \tag{18}$$

with $\chi^{-1} = \frac{\kappa}{2} - i\Delta$ the cavity's photonic or phononic response function, $\kappa = \kappa_{\mathrm{i}} + \kappa_{\mathrm{c}}$ the total decay rate, $\kappa_{\mathrm{i}} = \frac{\alpha' + \alpha L}{T}$ the intrinsic decay rate, $\kappa_{\mathrm{c}} = \frac{\mu}{T}$ the coupling rate, $\Delta = \frac{\varphi + \tilde{\Delta} L}{T}$ the detuning and $T = \frac{L}{v}$ the roundtrip time.

Next, we multiply (18) by $\sqrt{T}$ and switch from flux- to number-normalized fields ($\overline{a} \mapsto \sqrt{T}\overline{a}$):

$$\dot{\overline{a}} = v\sqrt{T}\,\overline{\zeta} - \chi^{-1}\overline{a} + \sqrt{\kappa_{\mathrm{c}}}s \tag{19}$$

From here on, $n = \overline{a}^\dagger\overline{a}$ represents the number of quanta in the cavity, while $s^\dagger s$ still corresponds to the injected photon or phonon flux. The transition from (14) to (19) still holds when we replace $z \mapsto -z$ because condition (16) also reverses. Therefore, potential dynamical differences between forward and backward scattering disappear in a high-finesse traveling-wave cavity – at least in the minimal model (2) of guided-wave optomechanics. Comparing (2) to (14), we see that $\zeta \propto fg$ with $f$ and $g$ equal to $a_{\mathrm{p/s}}$ or $b$. In the mean-field approximation, we assume these envelopes vary little over a roundtrip such that $\overline{fg} = \overline{f}\,\overline{g}$ holds (see Appendices). Finally, we apply the mean-field (14)-to-(19) transition to (2). Hence, an optomechanical cavity – constructed from a Brillouin-active waveguide – is governed by

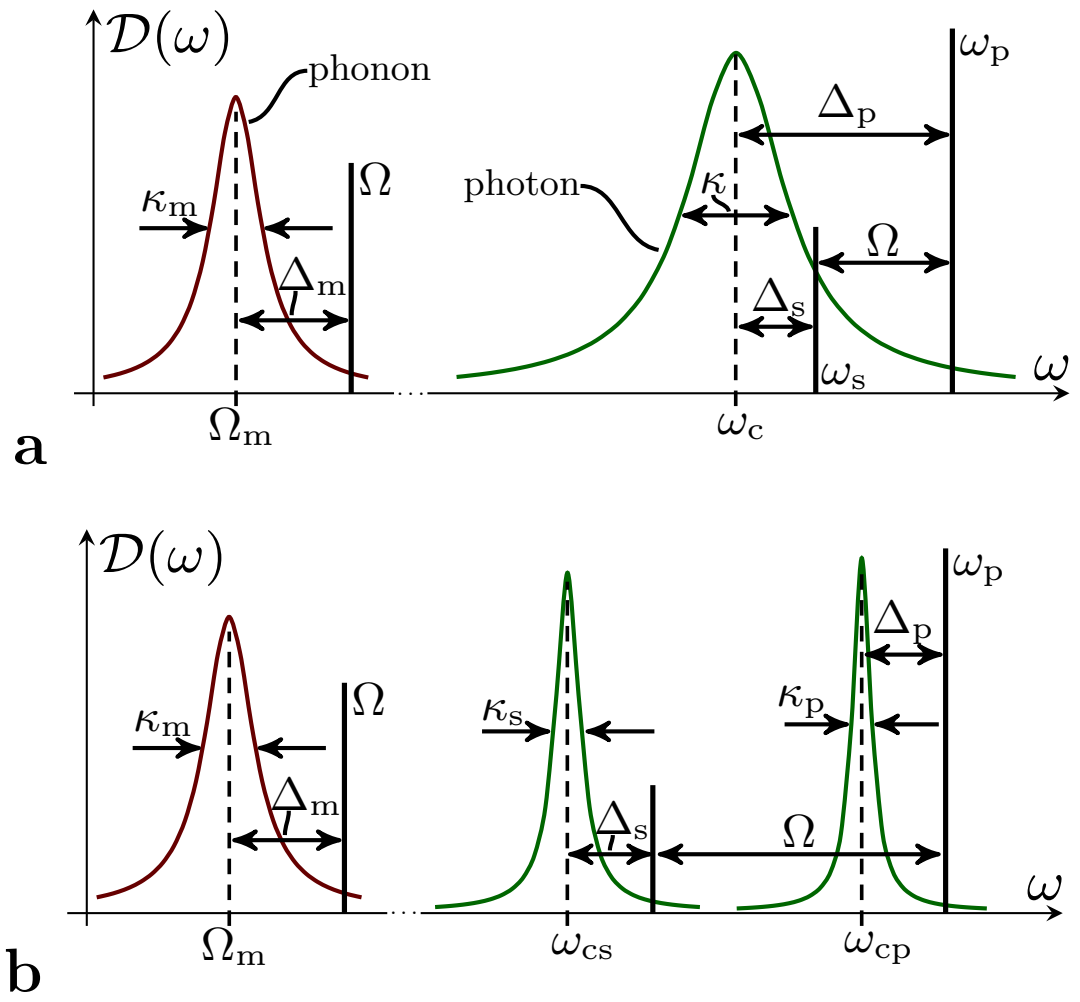

FIG. 5. **Cavity description.** The photonic and phononic density of states $\mathcal{D}(\omega)$. The mean-field equations (20) describe coupling between one phononic and either one (**a**) or two (**b**) photonic resonances. The latter case (**b**) is most power-efficient, although hard to achieve in practice [8].

$$\begin{aligned}
\dot{\overline{a}}_{\mathrm{p}} &= -ig_0\overline{a}_{\mathrm{s}}\overline{b} - \chi_{\mathrm{p}}^{-1}\overline{a}_{\mathrm{p}} + \sqrt{\kappa_{\mathrm{cp}}}s_{\mathrm{p}} \\
\dot{\overline{a}}_{\mathrm{s}} &= -ig_0\overline{b}^\dagger\overline{a}_{\mathrm{p}} - \chi_{\mathrm{s}}^{-1}\overline{a}_{\mathrm{s}} + \sqrt{\kappa_{\mathrm{cs}}}s_{\mathrm{s}} \\
\dot{\overline{b}} &= -ig_0\overline{a}_{\mathrm{s}}^\dagger\overline{a}_{\mathrm{p}} - \chi_{\mathrm{m}}^{-1}\overline{b} + \sqrt{\kappa_{\mathrm{cm}}}s_{\mathrm{m}}
\end{aligned} \tag{20}$$

with

$$\boxed{\sqrt{\frac{v_{\mathrm{p}}v_{\mathrm{s}}v_{\mathrm{m}}}{L}}\tilde{g}_0 = g_0} \tag{21}$$

the well-known temporal zero-point coupling rate [2]. Indeed, equations (20) are equivalent (see Appendices) to the Heisenberg equations of motion resulting from the well-known Hamiltonian $\hat{\mathcal{H}} = \hbar\omega_{\mathrm{c}}(\hat{x})\hat{a}^\dagger\hat{a} + \hbar\Omega_{\mathrm{m}}\hat{b}^\dagger\hat{b}$ [2]. Remarkably, the equivalence holds even for inter-modal and backward scattering. The connection (21) between the traveling-wave and the cavity-based vacuum coupling rates $\tilde{g}_0$ and $g_0$ is at the heart of this work: other links such as (1) are based on this result. Further, the mean-field transition transforms the guided-wave commutator

(3) into

$$\left[\bar{a}_\gamma, \bar{a}^\dagger_{\gamma'}\right] = \frac{\sqrt{v_\gamma v_{\gamma'}}}{L^2}\delta_{\gamma\gamma'}\int_0^L\int_0^L \mathrm{d}z\mathrm{d}z'\delta(z-z') = \frac{\sqrt{v_\gamma v_{\gamma'}}}{L}\delta_{\gamma\gamma'} \tag{22}$$

and through rescaling $\bar{a}_\gamma$ by $\sqrt{T_\gamma}$ into

$$\left[\bar{a}_\gamma, \bar{a}^\dagger_{\gamma'}\right] = \delta_{\gamma\gamma'} \tag{23}$$

thus correctly retrieving the standard harmonic oscillator commutators [2].

To derive (20), we made the same mean-field transition for photons and phonons. In particular, this supposes a large phonon finesse $\mathcal{F}_\mathrm{m} = \frac{2\pi}{\kappa_\mathrm{m} T_\mathrm{m}} \gg 1$. Often there is only intrinsic phonon loss such that $\kappa_\mathrm{m} = v_\mathrm{m}\alpha_\mathrm{m}$ and thus this requires $\frac{2\pi}{\alpha_\mathrm{m} L} \gg 1$. In many systems, the phonon decay length $\alpha_\mathrm{m}^{-1}$ is much shorter than the roundtrip length $L$. Then this phonon high-finesse limit does not hold. However, we can completely neglect phonon propagation ($\partial_z b \to 0$ in (2)) if $\alpha_\mathrm{m}$ is sufficiently large. The phonons' envelope operator $b$ then obeys

$$v_\mathrm{m}^{-1}\partial_t b = -i\tilde{g}_0 a_\mathrm{s}^\dagger a_\mathrm{p} - \tilde{\chi}_\mathrm{m}^{-1} b$$

Applying (13), multiplying by $\sqrt{T_\mathrm{m}}$ and switching from flux- to number-normalized envelopes results in

$$\dot{\bar{b}} = -ig_0\bar{a}_\mathrm{s}^\dagger\bar{a}_\mathrm{p} - v_\mathrm{m}\tilde{\chi}_\mathrm{m}^{-1}\bar{b} \tag{24}$$

where we used (21). Hence, this localized low-phonon-finesse approach yields the same result as the previous high-finesse limit with $s_\mathrm{m} = 0$ (see (20)). Therefore, even low-finesse phonons produce the same dynamics as is commonly studied in cavity optomechanics [2].

Notably, the standard treatment of cavity optomechanics [2] does not consider an explicit space variable: the Hamiltonian $\hat{\mathcal{H}}$ performs an implicit spatial average by describing the entire object as single mechanical oscillator, in contrast to the explicit spatial average (13) performed in this work. However, even the implicit average in $\hat{\mathcal{H}}$ requires low-loss acoustic excitations to set up a global mechanical mode self-consistently, precisely as in the high-finesse approximation leading to (20). In the localized, low-finesse phonon approach that generates (24), the spatial averaging can still be performed and yields the same classical dynamics – but its meaning changes. Now ($\mathcal{F}_\mathrm{m} < 1$) the acoustic wave is too lossy to set up a global mechanical mode for the entire cavity. Instead, the cavity consists of an ensemble of independent Raman-like mechanical oscillators. It is no longer possible to address phonons circulating in the cavity.

Finally, we combine (21) and (11). Using $\alpha_\mathrm{m} v_\mathrm{m} = \Omega_\mathrm{m}/Q_\mathrm{m}$, we obtain result (1) immediately. Note that $Q_\mathrm{m}$ is defined here as the waveguide's intrinsic phonon quality factor, which could be different from the cavity's phonon quality factor if there were e.g. non-negligible phonon coupling or bending losses. In case of doubt, it is safe to alternatively write (1) as

$$v_\mathrm{p} v_\mathrm{s} v_\mathrm{m}\frac{(\hbar\omega_\mathrm{p})\,\alpha_\mathrm{m}}{4L}\tilde{\mathcal{G}} = g_0^2 \tag{25}$$

Both $\tilde{\mathcal{G}}$ and $g_0$ are well-established in the study of photon-phonon interaction, but they operate on different levels. The Planck constant $\hbar$ enters (1) because $\tilde{\mathcal{G}}$ is classical while $g_0$ is inherently quantum mechanical. In addition, $\tilde{\mathcal{G}}$ quantifies the combined action of forces and scattering and contains the phonon loss – while $g_0$ does not. Further, larger $L$ yields a smaller $g_0$ while $\tilde{\mathcal{G}}$ is length-independent. Therefore, $g_0^2 \propto \frac{\hbar}{L}\frac{\tilde{\mathcal{G}}}{Q_\mathrm{m}}$. This mean-field derivation is but one way to prove the $\tilde{\mathcal{G}} \leftrightarrow g_0$ conversion, other approaches yield the same result (see Appendices). This proof captures all reversible photon-phonon coupling mechanisms.

## IV. SYMMETRY BETWEEN CIRCUIT AND CAVITY OPTOMECHANICAL EFFECTS

In this section, we describe both guided-wave and cavity-based regimes of photon-phonon coupling. To begin with, we recover and briefly review the known cavity-based regimes of photon lasing, phonon lasing and strong coupling. Next, we map these regimes on the guided-wave spatial evolution of the opto-acoustic fields. The mapping unveils two unobserved regimes of guided-wave Brillouin scattering. We pay particular attention to the strong coupling regime ($\tilde{g} \gg \alpha_\mathrm{s} + \alpha_\mathrm{m}$).

Here, we assume zero photon and phonon input flux and an undepleted pump. Then (20) reduces to

$$\begin{aligned}\dot{\bar{a}}_\mathrm{s} &= -ig_0\bar{b}^\dagger\bar{a}_\mathrm{p} - \chi_\mathrm{s}^{-1}\bar{a}_\mathrm{s} \\ \dot{\bar{b}} &= -ig_0\bar{a}_\mathrm{s}^\dagger\bar{a}_\mathrm{p} - \chi_\mathrm{m}^{-1}\bar{b}\end{aligned} \tag{26}$$

These equations treat the photons and phonons identically. Therefore, every photonic phenomenon must have a phononic counterpart and vice versa. Even more, the *temporal* cavity dynamics (26) can be mapped ($t \mapsto z$) on the *spatial* steady-state waveguide evolution (8). Each effect known from cavities therefore has a waveguide counterpart (but not vice versa as we will see). This also implies that the spatial figures of merit have a temporal symmetry partner and vice versa; we prepared for this at the end of section II by defining a guided-wave vacuum coupling rate $\tilde{g}_0$ and cooperativity $\tilde{\mathcal{C}}$. To clearly expose these symmetries, we solve (26); keeping in mind that the very same discussion holds spatially for (8). First, we decouple (26) and get

$$\left(\frac{\mathrm{d}}{\mathrm{d}t} + \chi_\mathrm{s}^{-\star}\right)\left(\frac{\mathrm{d}}{\mathrm{d}t} + \chi_\mathrm{m}^{-1}\right)\bar{b}(t) = g^2\bar{b}(t) \tag{27}$$

Here, we introduced the pump-enhanced coupling rate $g = g_0\sqrt{n_\mathrm{p}}$. Next, we insert the ansatz $\bar{b} \propto e^{\gamma t}$ in (27) and find two roots $\gamma_\pm$

$$\gamma_\pm = \frac{1}{2}\left\{-\left(\chi_{\rm s}^{-\star} + \chi_{\rm m}^{-1}\right) \pm \sqrt{\left(\chi_{\rm s}^{-\star} - \chi_{\rm m}^{-1}\right)^2 + 4g^2}\right\} \tag{28}$$

In general, these roots strongly mix the photon and phonon response: the photon-phonon pair forms a polariton [12, 56, 61, 68, 69]. The guided-wave analog of (27) is

$$\left(\partial_z \pm \tilde{\chi}_{\rm s}^{-\star}\right)\left(\partial_z + \tilde{\chi}_{\rm m}^{-1}\right) b(z) = \pm \tilde{g}^2 b(z) \tag{29}$$

and it can be treated identically. The full spatial and temporal dynamics is governed by the general solution (28) (see Appendices). However, it is more instructive to consider the limiting cases of weak and strong photon-phonon interaction relative to the system's damping.

First, if the photon-phonon interaction is sufficiently weak, i.e. $g \ll |\kappa_{\rm s} - \kappa_{\rm m}|$, the two roots in (28) disconnect. Usually, the photon and phonon decay rates differ significantly. Then there are two scenario's depending on the relative photonic and phononic decay rates. Essentially, the dynamics of the short-lived particle can be adiabatically eliminated, although it may still strongly modify the response of its long-lived partner.

In particular, when the phonons decay slowly ($\kappa_{\rm m} \ll \kappa_{\rm s}$), the photonic response is barely modified: $\dot{\bar{a}}_{\rm s} \to 0$ and therefore $\bar{a}_{\rm s} = -i\chi_{\rm s} g_0 \bar{b}^\dagger \bar{a}_{\rm p}$ to a good approximation. However, the phononic response can then dramatically change to $\chi_{\rm m}^{-1} - g^2\chi_{\rm s}^\star$. Hence, we recover the spring effect ($\delta\Omega_{\rm m} = g^2\Im\chi_{\rm s}^\star$) and phonon lasing ($\delta\kappa_{\rm m} = -2g^2\Re\chi_{\rm s}^\star$) [2]. At the photon resonance ($\Delta_{\rm s} = 0$), the phonon linewidth equals $\kappa_{\rm m} + \delta\kappa_{\rm m} = (1 - \mathcal{C})\,\kappa_{\rm m}$ with $\mathcal{C} = \frac{4g^2}{\kappa_{\rm s}\kappa_{\rm m}}$ the cooperativity. Therefore, the threshold for mechanical lasing is $\mathcal{C} = 1$. This instability was first contemplated by Braginsky [70] and received further study in systems ranging from gram-scale mirrors [71] to optomechanical crystals [72, 73]. Since $\langle n_{\rm s}\rangle = \frac{\kappa_{\rm m}}{\kappa_{\rm s}}\mathcal{C}\langle n_{\rm m}\rangle \ll \langle n_{\rm m}\rangle$, there are far fewer Stokes photons than phonons in the cavity in this situation. The system enters the strong coupling regime as soon as $\mathcal{C} \sim \frac{\kappa_{\rm s}}{\kappa_{\rm m}}$.

Similarly, when the photons decay slowly ($\kappa_{\rm s} \ll \kappa_{\rm m}$), the phononic response is barely modified: $\dot{\bar{b}} \to 0$ and therefore $\bar{b} = -i\chi_{\rm m} g_0 \bar{a}_{\rm s}^\dagger \bar{a}_{\rm p}$ to a good approximation. However, the photonic response can then dramatically change to $\chi_{\rm s}^{-1} - g^2\chi_{\rm m}^\star$. Hence, we recover the cavity frequency pull ($\delta\omega_{\rm cs} = g^2\Im\chi_{\rm m}^\star$) and photon lasing ($\delta\kappa_{\rm s} = -2g^2\Re\chi_{\rm m}^\star$) [4, 74, 75]. At the phonon resonance ($\Delta_{\rm m} = 0$), the Stokes linewidth equals $\kappa_{\rm s} + \delta\kappa_{\rm s} = (1 - \mathcal{C})\,\kappa_{\rm s}$ with $\mathcal{C}$ the same temporal cooperativity as above. Therefore, the threshold for Brillouin lasing is also $\mathcal{C} = 1$. First realized in fibers [76], this case was recently also studied in $CaF_2$ resonators [14], silica disks [8] and chalcogenide waveguides [15]. Such lasers are known for their excellent spectral purity [75, 77] and received attention for quantum-limited amplification [16]. Since $\langle n_{\rm m}\rangle = \frac{\kappa_{\rm s}}{\kappa_{\rm m}}\mathcal{C}\langle n_{\rm s}\rangle \ll \langle n_{\rm s}\rangle$, there are far fewer phonons than Stokes photons in the cavity in this situation. The system enters the strong coupling regime as soon as $\mathcal{C} \sim \frac{\kappa_{\rm m}}{\kappa_{\rm s}}$.

Further, if the photon-phonon coupling rate is sufficiently strong, (27) simplifies to $\ddot{\bar{b}} = g^2\bar{b}$. An identical derivation yields $\ddot{\bar{b}} = -g^2\bar{b}$ if the Stokes wave is considered undepleted. Therefore, a red-detuned probe produces entangled photon-phonon pair generation ($\bar{b}(t) \propto e^{\pm gt}$), whereas a blue-detuned probe produces Rabi flopping between photons and phonons ($\bar{b}(t) \propto e^{\pm igt}$) [2]. A situation of equally strong optical and mechanical damping ($\kappa_{\rm s} \approx \kappa_{\rm m}$) invalidates the above weak coupling treatment even for small $g$. However, this is not sufficient to see strong coupling behavior. From the general solution (see Appendices), this requires $g \gg \kappa_{\rm s} + \kappa_{\rm m}$. Indeed, in the strong coupling regime the hybridized photon-phonon polariton sees half the optical and half the mechanical damping [12]. Therefore, the state-swap frequency $\frac{g}{\pi}$ must be high compared to the average decay rate $\frac{\kappa_{\rm s}+\kappa_{\rm m}}{2}$ to observe an actual Rabi swap before the population decreases by $1/e$.

By comparing (26) and (8), we prove an analogy between spatial and temporal optomechanical effects (fig.3): the above cavity-based discussion still largely holds for guided-wave optomechanics with the mapping $g_0 \mapsto \tilde{g}_0$, $\mathcal{C} \mapsto \tilde{\mathcal{C}}$, $\kappa_{\rm s} \mapsto \alpha_{\rm s}$, $\kappa_{\rm m} \mapsto \alpha_{\rm m}$ and $n \mapsto \Phi$. In case of all co-propagating waves, and in the absence of cascading [30, 63] and noise [46, 47], the mapping of cavity optomechanics onto a Brillouin-active waveguide in steady-state is an exact equivalence. However, when for instance one of the particles counter-propagates, such as the Stokes photons in backward scattering, important differences arise that have no equivalent in cavity optomechanics. Indeed, as proven in section III, information regarding the propagation direction of the waves disappears in the mean-field transition. Instead comparing (27) and (29), much can still be learned by instead mapping $g_0^2 \mapsto -\tilde{g}_0^2$ and $\kappa_{\rm s} \mapsto -\alpha_{\rm s}$. Note that this particular difference disappears if the counter-propagating particle species is undepleted: then it vanishes from the dynamics and the situation is identical to the co-propagating case.

Thus, guided-wave weak coupling requires $\tilde{g} \ll |\alpha_{\rm s} \mp \alpha_{\rm m}|$ with $\tilde{g} = \tilde{g}_0\sqrt{\Phi_{\rm p}}$ the spatial coupling rate (see Appendices). Under weak coupling, there are two cases depending on the relative photon and phonon propagation losses. We have touched upon these subcases at the end of section II and briefly consider them again here to show the similarity with cavity-optomechanical effects. First, when the phonons propagate far ($\alpha_{\rm m} \ll \alpha_{\rm s}$), the photonic loss $\alpha_{\rm s}$ barely changes. However, the phononic response can then drastically change to $\tilde{\chi}_{\rm m}^{-1} - \tilde{g}^2\tilde{\chi}_{\rm s}^\star$, which includes a shift in both the phononic propagation loss ($\delta\alpha_{\rm m} = -2\tilde{g}^2\Re\tilde{\chi}_{\rm s}^\star$) and group velocity ($\propto \Im\tilde{\chi}_{\rm s}^\star$); i.e. traveling-phonon amplification and light-induced slowing down of sound. In section V, we show that this unobserved regime can be achieved in existing systems.

Second, when the Stokes photons propagate far ($\alpha_{\rm s} \ll \alpha_{\rm m}$), the phononic loss $\alpha_{\rm m}$ barely changes. However, the photonic response can then drastically change to $\tilde{\chi}_{\rm s}^{-1} - \tilde{g}^2\tilde{\chi}_{\rm m}^\star$. Hence, we are back in the conventional domain of phonon-assisted amplification of traveling photons ($\delta\alpha_{\rm s} = -2\tilde{g}^2\Re\tilde{\chi}_{\rm m}^\star$) and sound-induced slowing down of light ($\propto \Im\tilde{\chi}_{\rm m}^\star$) [38]. At resonance ($\tilde{\Delta}_{\rm m} = 0$), the Stokes propagation loss is $(1 - \tilde{\mathcal{C}})\alpha_{\rm s}$ as in (9).

If the coupling is sufficiently strong compared to the propagation losses ($\tilde{g} \gg \alpha_{\rm s}+\alpha_{\rm m}$), (29) simplifies to $\partial_z^2 b = \pm\tilde{g}^2 b$ (see Appendices). In the forward (+) case, and with boundary condition $b(0) = 0$, this yields

$$\begin{aligned} a_{\rm s}(z) &= a_{\rm s}(0)\cosh\tilde{g}z \\ b(z) &= -ia_{\rm s}^\dagger(0)\sinh\tilde{g}z \end{aligned} \tag{30}$$

such that $\Phi_{\rm s}(z) = \Phi_{\rm s}(0)\cosh^2\tilde{g}z$ and $\Phi_{\rm m}(z) = \Phi_{\rm s}(0)\sinh^2\tilde{g}z$. Therefore, $\Phi_{\rm s}(z) - \Phi_{\rm m}(z) = \Phi_s(0)$ and $\partial_z\Phi_{\rm s} = \partial_z\Phi_{\rm m}$ as required by the Manley-Rowe relations (4). In the backward (−) case, with $L$ the waveguide length and boundary condition $b(0) = 0$, the evolution along the waveguide has no cavity-optomechanics counterpart. Specifically, we retrieve

$$\begin{aligned} a_{\rm s}(z) &= \frac{a_{\rm s}(L)}{\cos\tilde{g}L}\cos\tilde{g}z \\ b(z) &= -i\frac{a_{\rm s}^\dagger(L)}{\cos\tilde{g}L}\sin\tilde{g}z \end{aligned} \tag{31}$$

such that $\Phi_{\rm s}(z) = \frac{\Phi_{\rm s}(L)}{\cos^2\tilde{g}L}\cos^2\tilde{g}z$ and $\Phi_{\rm m}(z) = \frac{\Phi_{\rm s}(L)}{\cos^2\tilde{g}L}\sin^2\tilde{g}z$. Therefore, $\Phi_{\rm s}(z) + \Phi_{\rm m}(z) = \frac{\Phi_{\rm s}(L)}{\cos^2\tilde{g}L}$ and $-\partial_z\Phi_{\rm s} = \partial_z\Phi_{\rm m}$ as required by Manley-Rowe (4). The system has an instability at $\tilde{g}L = \frac{\pi}{2}$, which is reached before a full state swap between light and sound can be completed. This situation is called "contraflow Hermitian coupling" in classifications of coupled-mode interactions [78, 79]. In case of anti-Stokes (instead of Stokes) seeding in the strong coupling regime, an identical derivation leads to $\partial_z^2 b = -\tilde{g}^2 b$ – which produces the same Rabi oscillations for forward and backward scattering. Although familiar in resonators [2], these strong-coupling effects have not yet been observed in the field of guided-wave Brillouin scattering; see section V for the prospects.

We conclude this section by analyzing the relation between the guided-wave and cavity-based cooperativities ($\tilde{\mathcal{C}}$ and $\mathcal{C}$) and by introducing a gain coefficient ($\mathcal{G}$) for an optomechanical cavity. Note that the temporal cooperativity

$$\mathcal{C} = \frac{4g^2}{\kappa_{\rm s}\kappa_{\rm m}} \tag{32}$$

is the ratio between the roundtrip gain and loss: inserting $g^2 = g_0^2 n_{\rm p}$, $n_{\rm p} = \frac{P_{\rm p}T_{\rm p}}{\hbar\omega_{\rm p}}$ and (21) in (32) indeed leads to

$$\mathcal{C} = \frac{\tilde{\mathcal{G}}P_{\rm p}}{\frac{\kappa_{\rm s}}{v_{\rm s}}}\frac{v_{\rm m}\alpha_{\rm m}}{\kappa_{\rm m}} = \frac{\tilde{\mathcal{G}}P_{\rm p}L}{\kappa_{\rm s}T_{\rm s}}\frac{v_{\rm m}\alpha_{\rm m}}{\kappa_{\rm m}} \tag{33}$$

with $P_{\rm p}$ the intracavity pump power and $\frac{v_{\rm m}\alpha_{\rm m}}{\kappa_{\rm m}}$ a naturally appearing correction factor that allows for higher phonon losses, so effectively lower $\mathcal{C}$, in the cavity than in the waveguide. This directly shows that

$$\tilde{\mathcal{C}} \geq \mathcal{C} \tag{34}$$

given (33), $\kappa_\gamma \geq v_\gamma\alpha_\gamma$ and $\tilde{\mathcal{C}} = \frac{\tilde{\mathcal{G}}P_{\rm p}}{\alpha_{\rm s}}$. Clearly, the guided-wave cooperativity exceeds the cavity-based cooperativity since the cavity has additional dissipation (e.g. coupling and bending losses). Finally, we define a gain coefficient $\mathcal{G}$ for a cavity in analogy to (11)

$$\mathcal{G} = \frac{4g_0^2}{\hbar\omega_{\rm p}\kappa_{\rm m}} \tag{35}$$

which characterizes the temporal exponential build-up of the Stokes when the phonons are heavily damped. The gain coefficients therefore obey

$$\tilde{\mathcal{G}} \geq \frac{L}{v_{\rm p}v_{\rm s}}\mathcal{G} \tag{36}$$

given $\kappa_{\rm m} \geq v_{\rm m}\alpha_{\rm m}$ and (21). Hence, the guided-wave and cavity-based optomechanical figures of merit are now conceptually similar and the relations between each of them were given in (1), (21), (34) and (36).

## V. PROSPECTS

In this section, we first give a couple of examples of how the $\tilde{\mathcal{G}} \leftrightarrow g_0$ connection (1) can be implemented – including several systems in which it can be tested empirically. Next, we move on to the prospects for observing new regimes of guided-wave optomechanics, simultaneously illustrating the application of our framework. Finally, we briefly discuss potential extensions to the minimal model (2) of traveling-wave Brillouin scattering.

Table I presents four implementations of the conversion from the gain coefficient $\tilde{\mathcal{G}}$ to the vacuum coupling rate $g_0$ ($\tilde{\mathcal{G}} \to g_0$) and four in reverse ($\tilde{\mathcal{G}} \leftarrow g_0$). The systems range from silicon nanowires and dual-web fibers to ultracold atom clouds and GaAs disks. In five cases, such as for silicon nanowires, the conversion can clearly be tested empirically by measuring $\tilde{\mathcal{G}}$ and $g_0$ through independent, established methods [2, 23]. In three cases, the conversion is hypothetical but still allows for comparison of the photon-phonon interaction strengths. For instance, an ultracold atom cloud in a Fabry-Pérot cavity [55] has no obvious traveling-wave equivalent. Nevertheless, its hypothetical waveguide partner would have a large gain coefficient of $\sim 10^8\ {\rm W}^{-1}{\rm m}^{-1}$ – which compares favorably to optomechanical waveguides realized to date.

So far, all Brillouin-active waveguides had far lower phonon than photon propagation lengths ($\alpha_{\rm m}^{-1} \ll \alpha_{\rm s}^{-1}$). Cavity-optomechanical systems, by contrast, more often than not had far lower phonon than photon damping rates ($\kappa_{\rm m} \ll \kappa_{\rm s}$) [2]. Only uniquely high-optical-quality [8, 14, 15, 74, 83, 84] systems succeed at reversing the latter hierarchy ($\kappa_{\rm m} \gg \kappa_{\rm s}$). The common reversal of this damping hierarchy (going from waveguides to cavities) stems from the small phonon group velocities ($v_\mathrm{m} \ll v_\mathrm{s}$).

| | $\tilde{\mathcal{G}}\,[\mathrm{W}^{-1}\mathrm{m}^{-1}]$ | $\underset{(1)}{\longleftrightarrow}$ | $\frac{g_0}{2\pi}$ [Hz] | $\frac{\Omega_\mathrm{m}}{2\pi}$ [Hz] | $Q_\mathrm{m}$ [−] | $L\,[\mu\mathrm{m}]$ | $n_\mathrm{g}$ [−] | $\lambda\,[\mu\mathrm{m}]$ |
|---|---|---|---|---|---|---|---|---|
| Silicon nanowire [23, 24] | $10^4$ | $\xrightarrow{\star}$ | $\frac{1.5\cdot 10^6}{\sqrt{L\,[\mu\mathrm{m}]}}$ | $10^{10}$ | $10^3$ | – | 4.6 | 1.55 |
| Silica standard fiber [4] | 1 | $\xrightarrow{\star}$ | $\frac{70}{\sqrt{L\,[\mathrm{cm}]}}$ | $10^{10}$ | 500 | – | 1.45 | 1.55 |
| Silica dual-web fiber [21] | $4\cdot 10^6$ | $\xrightarrow{\star}$ | $\frac{3\cdot 10^3}{\sqrt{L\,[\mathrm{cm}]}}$ | $6\cdot 10^6$ | $4\cdot 10^4$ | – | 1.7 | 1.55 |
| Chalcogenide rib [80, 81] | $3\cdot 10^2$ | $\xrightarrow{\star}$ | $\frac{7\cdot 10^5}{\sqrt{L\,[\mu\mathrm{m}]}}$ | $8\cdot 10^9$ | 230 | – | 2.6 | 1.55 |
| Silica microtoroid [12] | 600 | $\longleftarrow$ | $3\cdot 10^3$ | $8\cdot 10^7$ | $2\cdot 10^4$ | 97 | 1.45 | 0.78 |
| Silicon optomechanical crystal [73] | $4\cdot 10^4$ | $\xleftarrow{\star}$ | $6\cdot 10^5$ | $6\cdot 10^9$ | $2\cdot 10^3$ | 5 | 5 | 1.55 |
| Rb ultracold atom cloud [55] | $10^8$ | $\longleftarrow$ | $6\cdot 10^5$ | $4\cdot 10^4$ | 42 | 400 | 1 | 0.78 |
| GaAs optomechanical disk [82] | $5\cdot 10^4$ | $\longleftarrow$ | $3\cdot 10^5$ | $10^9$ | $2\cdot 10^3$ | 8 | 4 | 1.55 |

TABLE I. **Translation between waveguides and cavities.** The table contains four conversions from a gain coefficient to a vacuum coupling rate and four in reverse. The right five columns contain the parameters necessary for the conversion through formula (1). These are order-of-magnitude estimates. In some cases, indicated by a $\star$, the conversion can be empirically tested. In other situations, the conversion is hypothetical: e.g. an ultracold atom cloud in a Fabry-Pérot cavity has no obvious guided-wave equivalent.

The question naturally arises if waveguides with larger phonon than photon propagation length ($\alpha_\mathrm{m}^{-1} \gg \alpha_\mathrm{s}^{-1}$) can be made, while still keeping high cooperativities $\tilde{\mathcal{C}} = \frac{4\tilde{g}^2}{\alpha_\mathrm{s}\alpha_\mathrm{m}} \sim 1$. Currently, the largest phonon decay lengths are of the order $\alpha_\mathrm{m}^{-1} \sim 100\,\mu\mathrm{m}$ in backward Brillouin scattering [23, 65]. To realize larger phonon propagation lengths, one must look for waveguides with large acoustic group velocities $v_\mathrm{m}$ and small linewidths $\kappa_\mathrm{m}$. Thus, one promising approach uses low-frequency flexural modes ($\Omega \propto q^2$) in backward mode (large $q$) at low temperatures (large $Q_\mathrm{m}$). Indeed, then we have both large $v_\mathrm{m} \propto q$ and small $\kappa_\mathrm{m} = \frac{\Omega_\mathrm{m}}{Q_\mathrm{m}} \sim \frac{10^7}{10^4}\,\mathrm{Hz} = 1\,\mathrm{kHz}$ [12, 85, 86]. Since $\alpha_\mathrm{m}^{-1} = \frac{v_\mathrm{m}}{\kappa_\mathrm{m}}$ in a waveguide (where there is only propagation loss), we find that decay lengths up to $\alpha_\mathrm{m}^{-1} \sim 10\,\mathrm{m}$ are feasible given $v_\mathrm{m} \sim 10^4\,\mathrm{m/s}$ and $\kappa_\mathrm{m} \sim 1\,\mathrm{kHz}$. Such a small phonon propagation loss would strongly boost the cooperativity $\tilde{\mathcal{C}}$, which could compensate for a potentially lower coupling rate $\tilde{g}$ in backward mode. Clearly, nothing intrinsically forbids amplification of traveling phonons in systems such as the dual-web fiber [21] – where $\alpha_\mathrm{s}^{-1} \sim 10\,\mathrm{cm}$. Besides its scientific interest, such a traveling-phonon amplifier may be useful in phonon networks [13, 48, 49, 87, 88].

Next, we look into achieving the strong coupling regime in the typical situation of high acoustic loss ($\alpha_\mathrm{m} \gg \alpha_\mathrm{s}$). To see traveling-wave Rabi flopping, entangled photon-phonon pair production or contraflow Hermitian coupling (see section IV), one must obtain $\tilde{g} > \alpha_\mathrm{m}$ or equivalently $\tilde{\mathcal{C}} > \frac{\alpha_\mathrm{m}}{\alpha_\mathrm{s}}$. In optical fibers, in backward mode and given $\alpha_\mathrm{s}^{-1} \sim 10\,\mathrm{km}$ and $\alpha_\mathrm{m}^{-1} \sim 100\,\mu\mathrm{m}$, this requires $\tilde{\mathcal{C}} > 10^8$. This necessitates an unrealistic continuous-wave pump power of $P_\mathrm{p} > 10^8 \frac{\alpha_\mathrm{s}}{\tilde{\mathcal{G}}} = 10\,\mathrm{kW}$ with $\tilde{\mathcal{G}} \sim 1\,\mathrm{W}^{-1}\mathrm{m}^{-1}$.

In contrast, silicon chips can produce significantly lower $\frac{\alpha_\mathrm{m}}{\alpha_\mathrm{s}}$ ratios and therefore ease the condition on $\tilde{\mathcal{C}}$ for strong coupling. One can expect phonon propagation distances up to $\alpha_\mathrm{m}^{-1} \sim 1\,\mathrm{mm}$, as these are readily achieved in surface-acoustic-wave devices [89]. Together with $\alpha_\mathrm{s}^{-1} \sim 1\,\mathrm{cm}$ [23], this yields $\tilde{\mathcal{C}} > 10$ as the strong coupling condition, which requires a reasonable pump power of $P_\mathrm{p} > 10\frac{\alpha_\mathrm{s}}{\tilde{\mathcal{G}}} = 100\,\mathrm{mW}$ with $\tilde{\mathcal{G}} \sim 10^4\,\mathrm{W}^{-1}\mathrm{m}^{-1}$ [23, 24]. Indeed, current nanoscale silicon systems have already demonstrated $\tilde{\mathcal{C}} \approx 2$ [24, 63]. Hence, taking into account the rapid progress in state-of-the-art devices [1, 23, 24, 63], we expect demonstrations of traveling-phonon amplification and guided-wave strong coupling in the coming years. Such observations would open up entirely new realms of optomechanics.

Finally, this work can be extended on several fronts. First, the mean-field transition can be applied to the noise models of [46, 47]. Second, the regime of "nonlinear quantum optomechanics" [2, 90–92] should be transferred to waveguides. This requires that strong coupling is reached for merely one pump photon ($\Phi_\mathrm{p} \mapsto 1\,\mathrm{s}^{-1}$): $\tilde{g} = \tilde{g}_0\sqrt{\Phi_\mathrm{p}} = \tilde{g}_0 > \alpha_\mathrm{s} + \alpha_\mathrm{m}$. As $\alpha_\mathrm{m} \gg \alpha_\mathrm{s}$ usually, traveling-wave nonlinear quantum optomechanics is achieved when $\tilde{g}_0 > \alpha_\mathrm{m}$. Third, the coupling between the phononic mode and the thermal bath [46, 47] must be treated carefully to obtain truly quantum-coherent [12] coupling. And fourth, we focused mainly on the dynamics that optomechanical waveguides and cavities have in common, but wisdom may be found in the differences as well. We gave the example of contraflow strong coupling in section IV. In addition, the cavity has input fluxes that have no equivalent in a typical guided-wave set-up, while the waveguide can display spatiotemporal effects (both $\partial_z$ and $\partial_t$ in (2)) that are absent in a cavity (only $\partial_t$ in (20)). On top of this, the cavity breaks the symme-try between Stokes and anti-Stokes scattering, whereas this symmetry prevents exponential build-up of noise in low-dispersion forward intra-modal scattering [47]. It has also yet to be determined whether different cavity dynamics results in the medium-finesse case, as section III was limited to a low or high finesse.

With slight modifications, (2) also captures Raman effects [4, 6, 93]. For instance, the phonon frequency is much larger so an optical phase-mismatch can arise. Further, the thousandfold higher optical phonon frequency puts most Raman modes in the ground state at room temperature. This breaks the symmetry between Stokes and anti-Stokes scattering, such that thermally seeded exponential build-up may be seen even in the forward intra-modal case [47]. Some parameters introduced here – such as the spatial vacuum coupling $\tilde{g}_0$ – lose elegance in the Raman case: the impossibility of significant optical phonon transport undermines their symmetric definitions. Mathematically, expressions such as (5) diverge as $v_\mathrm{m}$ and $\alpha_\mathrm{m}^{-1}$ vanish. Then one must resort to the broken-symmetry Raman gain coefficient, which nevertheless obeys (1). Therefore, the core of this work also applies to guided-wave [4, 94–96], cavity-based [97–100] and surface-enhanced Raman scattering [57].

## VI. CONCLUSION

In conclusion, we unveiled a connection between Brillouin-active waveguides and optomechanical cavities. The link between the Brillouin gain coefficient $\tilde{\mathcal{G}}$ and the zero-point coupling rate $g_0$ was derived in a platform-independent way. As illustrated for silicon nanowires and ultracold atom clouds, it significantly expands the variety of systems whose photon-phonon interaction efficiency can be compared. Through the mean-field transition, we derived the dynamics of optomechanical cavities from that of Brillouin-active waveguides. We framed the behavior of both systems in terms of cooperativities and vacuum coupling rates. Next, we showed that phenomena familiar from cavity optomechanics all have guided-wave partners, but not the other way around. The broader theory predicts that several novel regimes, such as guided-wave strong coupling, will be accessible in state-of-the-art systems in the coming years. Hence, we showed that Brillouin scattering and cavity optomechanics are subsets of a larger realm of photon-phonon interaction.

## ACKNOWLEDGEMENT

R.V.L. acknowledges the Agency for Innovation by Science and Technology in Flanders (IWT) for a PhD grant. This work was partially funded under the FP7-ERC-InSpectra programme and the ITN-network cQOM. R.V.L. thanks T.J. Kippenberg for related discussions.

## Appendix A: Link to cavity Hamiltonian

With the mean-field transition derived in the main text, we take a step beyond the $\tilde{\mathcal{G}} \leftrightarrow g_0$ link. As we show in this section, the mean-field equations are equivalent to the cavity Langevin equations in the resolved-sideband limit ($\kappa \ll \Omega_{\mathrm{m}}$). In the case of coupling between one mechanical and one optical resonance (fig.2a), the usual theory [2] starts from the Hamiltonian

$$\hat{\mathcal{H}} = \hbar\omega_{\mathrm{c}}\hat{a}^\dagger\hat{a} + \hbar\Omega_{\mathrm{m}}\hat{b}^\dagger\hat{b} + \hat{\mathcal{H}}_{\mathrm{int}}$$

with

$$\hat{\mathcal{H}}_{\mathrm{int}} = \hbar g_0 \hat{a}^\dagger\hat{a}\left(\hat{b} + \hat{b}^\dagger\right)$$

the interaction Hamiltonian, $\hat{x} = x_{\mathrm{ZPF}}\left(\hat{b} + \hat{b}^\dagger\right)$ the mechanical oscillator's position, $x_{\mathrm{ZPF}}$ the zero-point motion, $\hat{a}$ and $\hat{b}$ ladder operators for the optical and mechanical oscillator and $g_0 = x_{\mathrm{ZPF}}\frac{\partial\omega_{\mathrm{c}}}{\partial x}$ the zero-point coupling rate. When the pump is undepleted, the interaction Hamiltonian can be linearized: $\hat{a} = \overline{\alpha} + \delta\hat{a}$ with $\delta\hat{a}$ a small fluctuation. Then we have

$$\hat{\mathcal{H}}_{\mathrm{int}}^{(\mathrm{lin})} = \hbar g_0\overline{\alpha}\left(\delta\hat{a} + \delta\hat{a}^\dagger\right)\left(\hat{b} + \hat{b}^\dagger\right)$$

Using the equation of motion $\dot{\hat{a}} = -\frac{i}{\hbar}[\hat{a}, \hat{\mathcal{H}}]$ and the commutator $[\hat{a}, \hat{a}^\dagger] = 1$ (the same for $\hat{b}$), this linearized Hamiltonian leads straight to the coupled equations [2]

$$\begin{aligned}
\delta\dot{\hat{a}} &= -\left(\frac{\kappa}{2} - i\Delta\right)\delta\hat{a} - ig_0\overline{\alpha}\left(\hat{b} + \hat{b}^\dagger\right) \\
\dot{\hat{b}} &= -\left(\frac{\kappa_{\mathrm{m}}}{2} - i\Omega_{\mathrm{m}}\right)\hat{b} - ig_0\overline{\alpha}\left(\delta\hat{a} + \delta\hat{a}^\dagger\right)
\end{aligned}$$

with $\Delta = \omega_{\mathrm{p}} - \omega_{\mathrm{c}}$. Next, we consider a blue-detuned pump in the resolved-sideband regime ($\kappa \ll \Omega_{\mathrm{m}}$). Then we can write the ladder operators as $\delta\hat{a} \to \hat{a}_{\mathrm{s}}e^{i\Omega t}$ and $\hat{b} \to \hat{b}e^{-i\Omega t}$, with $\hat{a}_{\mathrm{s}}$ and $\hat{b}$ now slowly-varying. We neglect the $\hat{b}$-term in the optical equation and the $\delta\hat{a}$-term in the mechanical equation because they are off-resonant. This is the rotating-wave approximation, which corresponds to the classical slowly-varying envelope approximation [3, 4]. Hence, the above equations reduce to

$$\begin{aligned}
\dot{\hat{a}}_{\mathrm{s}} &= -ig_0\overline{\alpha}\hat{b}^\dagger - \chi_{\mathrm{s}}^{-1}\hat{a}_{\mathrm{s}} \\
\dot{\hat{b}} &= -ig_0\overline{\alpha}\hat{a}_{\mathrm{s}}^\dagger - \chi_{\mathrm{m}}^{-1}\hat{b}
\end{aligned} \tag{A1}$$

and we find that equations (A1) are identical to equations (16) given $\hat{a}_{\mathrm{s}} \mapsto \overline{a}_{\mathrm{s}}$ and $\hat{b} \mapsto \overline{b}$. Remarkably, the equivalence holds even though the pump and Stokes could be counter-propagating or in different optical modes. In the unresolved-sideband limit ($\Omega_{\mathrm{m}} \ll \kappa$), anti-Stokes generation and cascading must be added for forward intra-modal, but not necessarily for backward or inter-modal Brillouin scattering. Indeed, comb generation is usually not accessible by backward or inter-modal coupling because of the phase-mismatch (fig.2). This assumption can be violated in Fabry-Pérot cavities [101] or when the first-order Stokes becomes sufficiently strong to pump a second-order Stokes wave [9].

## Appendix B: Manley-Rowe relations

In this section, we prove that the Manley-Rowe relations guarantee the existence of a single real, positive photon-phonon coupling coefficient in waveguides ($\tilde{g}_0$) and in cavities ($g_0$). In waveguides, the Manley-Rowe relations are formulated at the level of photon and phonon fluxes $\Phi$. In cavities, they are written down in terms of the total photon and phonon numbers $n$.

### *a. Manley-Rowe in waveguides*

A Brillouin-active waveguide in steady-state ($\partial_t \to 0$) obeys (see (2))

$$\begin{aligned}
\partial_z a_{\mathrm{p}} &= -i\tilde{\kappa}_{\mathrm{mop}}a_{\mathrm{s}}b - \tilde{\chi}_{\mathrm{p}}^{-1}a_{\mathrm{p}} \\
\pm\partial_z a_{\mathrm{s}} &= -i\tilde{\kappa}_{\mathrm{mos}}b^\dagger a_{\mathrm{p}} - \tilde{\chi}_{\mathrm{s}}^{-1}a_{\mathrm{s}} \\
\partial_z b &= -i\tilde{\kappa}_{\mathrm{om}}a_{\mathrm{s}}^\dagger a_{\mathrm{p}} - \tilde{\chi}_{\mathrm{m}}^{-1}b
\end{aligned} \tag{B1}$$

with arbitrary normalizations of the pump, Stokes and acoustic envelope such that generally $\tilde{\kappa}_{\mathrm{mop}} \neq \tilde{\kappa}_{\mathrm{mos}} \neq \tilde{\kappa}_{\mathrm{om}}$ are different complex numbers. Using $\partial_z\left(a^\dagger a\right) = (\partial_z a^\dagger)a + a^\dagger(\partial_z a)$, we find

$$\begin{aligned}
\partial_z\Phi_{\mathrm{p}} &= -\alpha_{\mathrm{p}}\Phi_{\mathrm{p}} + \left(i\tilde{\kappa}_{\mathrm{mop}}^\star a_{\mathrm{s}}^\dagger b^\dagger a_{\mathrm{p}} + \mathrm{h.c.}\right) \\
\pm\partial_z\Phi_{\mathrm{s}} &= -\alpha_{\mathrm{s}}\Phi_{\mathrm{s}} - \left(i\tilde{\kappa}_{\mathrm{mos}}a_{\mathrm{s}}^\dagger b^\dagger a_{\mathrm{p}} + \mathrm{h.c.}\right) \\
\partial_z\Phi_{\mathrm{m}} &= -\alpha_{\mathrm{m}}\Phi_{\mathrm{m}} - \left(i\tilde{\kappa}_{\mathrm{om}}a_{\mathrm{s}}^\dagger b^\dagger a_{\mathrm{p}} + \mathrm{h.c.}\right)
\end{aligned} \tag{B2}$$

Suppose now that the envelopes are flux-normalized such that $\Phi_{\mathrm{p}} = a_{\mathrm{p}}^\dagger a_{\mathrm{p}}$, $\Phi_{\mathrm{s}} = a_{\mathrm{s}}^\dagger a_{\mathrm{s}}$ and $\Phi_{\mathrm{m}} = b^\dagger b$ correspond to the number of pump photons, Stokes photons and phonons passing through a cross-section of the waveguide per second. Then we demand that, in the lossless case ($\alpha_{\mathrm{j}} = 0$), the rate of pump photon destruction equals the rate of Stokes photon and phonon creation

$$-\partial_z\Phi_{\mathrm{p}} = \pm\partial_z\Phi_{\mathrm{s}} = \partial_z\Phi_{\mathrm{m}} \tag{B3}$$

These are the Manley-Rowe relations [3, 79] for a Brillouin waveguide. We deduce from (B2) and (B3) that

$$\tilde{\kappa}_{\mathrm{mop}}^\star = \tilde{\kappa}_{\mathrm{mos}} = \tilde{\kappa}_{\mathrm{om}} \tag{B4}$$

This proves the existence of a single coupling coefficient that captures all reversible optical forces and scattering. Note that (B4) also guarantees power-conservation since

$$\partial_z\left(\hbar\omega_{\mathrm{p}}\Phi_{\mathrm{p}} \pm \hbar\omega_{\mathrm{s}}\Phi_{\mathrm{s}} + \hbar\Omega\Phi_{\mathrm{m}}\right) = 0$$

leads with (B2) in the lossless case to

$$-\omega_{\mathrm{p}}\tilde{\kappa}_{\mathrm{mop}}^\star + \omega_{\mathrm{s}}\tilde{\kappa}_{\mathrm{mos}} + \Omega\tilde{\kappa}_{\mathrm{om}} = 0 \tag{B5}$$

which is true given (B4) and $\omega_{\mathrm{p}} = \omega_{\mathrm{s}} + \Omega$. Next, we show that this coefficient (B4) can be taken real and positive without loss of generality. Renormalizing the envelopes to $c_{\mathrm{p}}a_{\mathrm{p}}$, $c_{\mathrm{s}}a_{\mathrm{s}}$ and $c_{\mathrm{m}}b$ yields new coupling coefficients

$$\frac{c_{\mathrm{p}}}{c_{\mathrm{s}}c_{\mathrm{m}}}\tilde{\kappa}_{\mathrm{mop}} \qquad \frac{c_{\mathrm{s}}}{c_{\mathrm{p}}c_{\mathrm{m}}^\star}\tilde{\kappa}_{\mathrm{mos}} \qquad \frac{c_{\mathrm{m}}}{c_{\mathrm{p}}c_{\mathrm{s}}^\star}\tilde{\kappa}_{\mathrm{om}} \tag{B6}$$

as can be seen from (B1). Suppose that $\tilde{\kappa}_{\rm om} = \tilde{g}_0 e^{i\varphi}$ is complex with $\tilde{g}_0$ real and positive. Then we take $c_{\rm p} = c_{\rm s} = c_{\rm m} = e^{-i\varphi}$. Using (B4) and (B6), it follows that the renormalized coupling coefficients are real and positive:

$$\tilde{\kappa}_{\rm mop} = \tilde{\kappa}_{\rm mos} = \tilde{\kappa}_{\rm om} = \tilde{g}_0 \tag{B7}$$

This unique coupling coefficient quantifies the coupling strength between a single photon and a single phonon propagating along a waveguide. Indeed, suppose that $a_{\rm p} = a_{\rm s} = b \mapsto 1\,{\rm s}^{-1/2}$ such that $\Phi_{\rm p} = \Phi_{\rm s} = \Phi_{\rm m} \mapsto 1\,{\rm s}^{-1}$ at a certain point along the waveguide. In the lossless case, (B2) then becomes

$$\begin{aligned} \partial_z \Phi_{\rm p} &= -2\tilde{g}_0 \\ \pm\partial_z \Phi_{\rm s} &= 2\tilde{g}_0 \\ \partial_z \Phi_{\rm m} &= 2\tilde{g}_0 \end{aligned} \tag{B8}$$

So $2\tilde{g}_0$ gives the rate (per meter) at which the pump flux decreases and the Stokes and phonon flux increase at a point along waveguide through which one pump photon, one Stokes photon and one phonon are passing.

The waveguide coupling coefficient $\tilde{g}_0$ can also be interpreted in terms of a zero-point motion. As shown in (14), we have

$$\tilde{g}_0 = \sqrt{\frac{L}{v_{\rm p} v_{\rm s} v_{\rm m}}}\, g_0 \tag{B9}$$

For forward intra-modal scattering ($v_{\rm p} = v_{\rm s} = v_{\rm g}$)

$$g_0 = x_{\rm ZPF} \left.\frac{\partial \omega_{\rm p}}{\partial x}\right|_{k_{\rm p}} \tag{B10}$$

is defined in terms of the zero-point motion and the cavity frequency pull at fixed wavevector [2]. Combining (B9), (B10) and (D9), we obtain

$$\tilde{g}_0 = -\frac{\omega_{\rm p}}{c} \tilde{x}_{\rm ZPF} \left.\frac{\partial n_{\rm eff}}{\partial x}\right|_{\omega_{\rm p}} = -\tilde{x}_{\rm ZPF} \left.\frac{\partial k_{\rm p}}{\partial x}\right|_{\omega_{\rm p}} \tag{B11}$$

with

$$\tilde{x}_{\rm ZPF} = x_{\rm ZPF} \sqrt{\frac{L}{v_{\rm m}}} = \sqrt{\frac{\hbar}{2 m_{\rm eff} v_{\rm m} \Omega_{\rm m}}} \tag{B12}$$

the waveguide "zero-point motion" and $m_{\rm eff}$ the effective mass per unit length. Indeed, a waveguide section of length $L$ contains on average $\langle n_{\rm m} \rangle = \frac{L}{v_{\rm m}} \langle \Phi_{\rm m} \rangle$ phonons with $\langle \Phi_{\rm m} \rangle$ the mean phonon flux. As fluxes – instead of numbers – are the fundamental quantities in waveguides, the zero-point motion is corrected by precisely a factor $\sqrt{\frac{L}{v_{\rm m}}}$ in (B12).

Often the optical envelopes are power-normalized and the acoustic envelope displacement-normalized. Starting from flux-normalized envelopes, one can switch to such normalizations through

$$c_{\rm p} = \sqrt{\hbar \omega_{\rm p}} \qquad c_{\rm s} = \sqrt{\hbar \omega_{\rm s}} \qquad c_{\rm m} = \sqrt{\frac{2\hbar \Omega_{\rm m}}{k_{\rm eff} v_{\rm m}}} = 2\tilde{x}_{\rm ZPF} \tag{B13}$$

with $k_{\rm eff}$ the effective stiffness per unit length and by applying (B6).

*b. Manley-Rowe in cavities*

Here, we apply the discussion of the previous section to the mean-field cavity equations. With arbitrary envelope normalizations and without input, equations (13) are

$$\begin{aligned} \dot{\overline{a}}_{\rm p} &= -i\kappa_{\rm mop} \overline{a}_{\rm s} \overline{b} - \chi_{\rm p}^{-1} \overline{a}_{\rm p} \\ \dot{\overline{a}}_{\rm s} &= -i\kappa_{\rm mos} \overline{b}^\dagger \overline{a}_{\rm p} - \chi_{\rm s}^{-1} \overline{a}_{\rm s} \\ \dot{\overline{b}} &= -i\kappa_{\rm om} \overline{a}_{\rm s}^\dagger \overline{a}_{\rm p} - \chi_{\rm m}^{-1} \overline{b} \end{aligned} \tag{B14}$$

with generally $\kappa_{\rm mop} \neq \kappa_{\rm mos} \neq \kappa_{\rm om}$. Applying $\frac{\rm d}{dt}\left(\overline{a}^\dagger \overline{a}\right) = (\frac{\rm d}{{\rm d}t}\overline{a}^\dagger)\overline{a} + \overline{a}^\dagger(\frac{\rm d}{{\rm d}t}\overline{a})$ to (B14), we find

$$\begin{aligned} \frac{\rm d}{{\rm d}t} n_{\rm p} &= -\kappa_{\rm p} n_{\rm p} + \left(i\kappa_{\rm mop}^\star \overline{a}_{\rm s}^\dagger \overline{b}^\dagger \overline{a}_{\rm p} + \text{h.c.}\right) \\ \frac{\rm d}{{\rm d}t} n_{\rm s} &= -\kappa_{\rm s} n_{\rm s} - \left(i\kappa_{\rm mos} \overline{a}_{\rm s}^\dagger \overline{b}^\dagger \overline{a}_{\rm p} + \text{h.c.}\right) \\ \frac{\rm d}{{\rm d}t} n_{\rm m} &= -\kappa_{\rm m} n_{\rm m} - \left(i\kappa_{\rm om} \overline{a}_{\rm s}^\dagger \overline{b}^\dagger \overline{a}_{\rm p} + \text{h.c.}\right) \end{aligned} \tag{B15}$$

Suppose now that the envelopes are number-normalized such that $n_{\rm p} = \overline{a}_{\rm p}^\dagger \overline{a}_{\rm p}$, $n_{\rm s} = \overline{a}_{\rm s}^\dagger \overline{a}_{\rm s}$ and $n_{\rm m} = \overline{b}^\dagger \overline{b}$ correspond to the number of pump photons, Stokes photons and phonons in the cavity. We demand that, in the lossless case ($\kappa_{\rm j} = 0$), the rate of pump photon destruction equals the rate of Stokes photon and phonon creation

$$-\dot{n}_{\rm p} = \dot{n}_{\rm s} = \dot{n}_{\rm m} \tag{B16}$$

These are the Manley-Rowe equations for an optomechanical cavity. We deduce from (B15) and (B16) that

$$\kappa_{\rm mop}^\star = \kappa_{\rm mos} = \kappa_{\rm om} \tag{B17}$$

This proves the existence of a single coupling coefficient that captures all conservative optical forces and scattering. Note that (B17) also guarantees energy-conservation since

$$\frac{\rm d}{{\rm d}t}\left(\hbar\omega_{\rm p} n_{\rm p} + \hbar\omega_{\rm s} n_{\rm s} + \hbar\Omega n_{\rm m}\right) = 0$$

leads with (B15) in the lossless case to

$$-\omega_{\rm p}\kappa_{\rm mop}^\star + \omega_{\rm s}\kappa_{\rm mos} + \Omega\kappa_{\rm om} = 0 \tag{B18}$$

which holds given (B17) and $\omega_{\rm p} = \omega_{\rm s} + \Omega$. As in the previous section, one can show that this coupling coefficient can be chosen real and positive. This unique coupling coefficient must then be the well-known $g_0$. It quantifies the interaction strength between a single photon and a single phonon trapped in a cavity. Indeed, suppose that $\overline{a}_{\rm p} = \overline{a}_{\rm s} = \overline{b} \mapsto 1$ such that $n_{\rm p} = n_{\rm s} = n_{\rm m} \mapsto 1$ at a certain point in time. In the lossless case, (B15) then becomes

$$\begin{aligned} \dot{n}_{\rm p} &= -2g_0 \\ \dot{n}_{\rm s} &= 2g_0 \\ \dot{n}_{\rm m} &= 2g_0 \end{aligned} \tag{B19}$$

So $2g_0$ gives the rate (per second) at which the number of pump photons decreases and the number of Stokes photons and phonons increases when there is one pump photon, one Stokes photon and one phonon in the cavity.

Often the optical envelopes are energy-normalized and the acoustic envelope displacement-normalized. Starting from number-normalized envelopes, one can switch to such normalizations through

$$c_{\mathrm{p}} = \sqrt{\hbar\omega_{\mathrm{p}}} \qquad c_{\mathrm{s}} = \sqrt{\hbar\omega_{\mathrm{s}}} \qquad c_{\mathrm{m}} = \sqrt{\frac{2\hbar\Omega_{\mathrm{m}}}{k_{\mathrm{eff}}L}} = 2x_{\mathrm{ZPF}} \tag{B20}$$

with $x_{\mathrm{ZPF}}$ the zero-point motion and by applying (B6).

## Appendix C: Mean-field approximation

### c. *Justification of $\overline{fg} = \overline{f}\overline{g}$*

We denote $f(z,t)$ and $g(z,t)$ two operators that vary slowly on a lengthscale $L$. The mean-field operators are defined as $\overline{f}(t) = \frac{1}{L}\int_0^L f(z,t)\mathrm{d}z$. Clearly, when $f(z,t) = f(0,t)$ and $g(z,t) = g(0,t)$ are constants then $\overline{fg}(t) = f(0,t)g(0,t) = \overline{f}(t)\overline{g}(t)$. Let us assume now that the amplitudes vary slowly enough such that they can be Taylor-expanded as $f(z,t) = f(0,t) + f'z$ with $f' = \partial_z f(0,t)$ and the same for $g$. Then we have

$$\overline{f} = \frac{1}{L}\left(f(0)L + f'(0)\frac{L^2}{2}\right)$$
$$\overline{g} = \frac{1}{L}\left(g(0)L + g'(0)\frac{L^2}{2}\right)$$

where we dropped the time-dependence. Thus, we have

$$\overline{f}\overline{g} = f(0)g(0) + (f'g(0) + f(0)g')\,\frac{L}{2} + f'g'\frac{L^2}{4}$$

Similarly,

$$\begin{aligned}\overline{fg} &= \frac{1}{L}\int_0^L \left(f(0)g(0) + (f'g(0) + f(0)g')\,z + f'g'z^2\right)\mathrm{d}z \\ &= f(0)g(0) + (f'g(0) + f(0)g')\,\frac{L}{2} + f'g'\frac{L^2}{3}\end{aligned}$$

Therefore $\overline{fg} - \overline{f}\overline{g} = f'g'\frac{L^2}{12}$ which can be neglected if $L$ is sufficiently small compared to the lengthscale on which $f(z,t)$ and $g(z,t)$ vary.

## Appendix D: Alternative proofs of the $\tilde{\mathcal{G}} \leftrightarrow g_0$ link

In this section, we describe two other approaches to derive the link

$$g_0^2 = v_{\mathrm{g}}^2 \frac{(\hbar\omega_p)\,\Omega_{\mathrm{m}}}{4L}\left(\frac{\tilde{\mathcal{G}}}{Q_{\mathrm{m}}}\right) \tag{D1}$$

### d. *From independent full-vectorial definitions*

Here, we derive equation (D1) from the full-vectorial definitions of $\tilde{\mathcal{G}}$ and $g_0$ – specializing to intra-modal forward scattering. We focus on the moving boundary contribution. From the perturbation theory of Maxwell's equations with respect to moving boundaries [102], the cavity frequency shift $\frac{\partial\omega_{\mathrm{c}}}{\partial x}$ can be expressed as

$$\frac{\partial\omega_{\mathrm{c}}}{\partial x} = \frac{\omega_{\mathrm{p}}}{2}\frac{\oint dA\,(\mathbf{u}\cdot\hat{\mathbf{n}})\left(\Delta\epsilon|\mathbf{E}_{\|}|^2 - \Delta\epsilon^{-1}|\mathbf{D}_{\perp}|^2\right)}{\int dV\epsilon|\mathbf{E}|^2}$$

with $\mathbf{u}$ the normalized ($\max(|\mathbf{u}|) = 1$) acoustic field, $\hat{\mathbf{n}}$ the unit normal pointing from material 1 to material 2, $\Delta\epsilon = \epsilon_1 - \epsilon_2$ and $\Delta\epsilon^{-1} = \epsilon_1^{-1} - \epsilon_2^{-1}$. The upper integral is over the entire surface area of the cavity, the lower integral across the cavity volume. Further, $\mathbf{E}_{\|}$ is the electric field parallel to the boundary and $\mathbf{D}_{\perp}$ the displacement field perpendicular to the boundary. For a longitudinally invariant cavity, the surface integral can be reduced to a line integral and the volume integral to a surface integral:

$$\frac{\partial\omega_{\mathrm{c}}}{\partial x} = \frac{\omega_{\mathrm{p}}}{2}\frac{\oint dl\,(\mathbf{u}\cdot\hat{\mathbf{n}})\left(\Delta\epsilon|\mathbf{E}_{\|}|^2 - \Delta\epsilon^{-1}|\mathbf{D}_{\perp}|^2\right)}{\int dA\epsilon|\mathbf{E}|^2} \tag{D2}$$

Further, the gain coefficient $\tilde{\mathcal{G}}$ is given by [23, 25, 103]

$$\tilde{\mathcal{G}} = \omega_{\mathrm{p}}\frac{Q_{\mathrm{m}}}{2k_{\mathrm{eff}}}|\langle\mathbf{f},\mathbf{u}\rangle|^2 \tag{D3}$$

with $\mathbf{f}$ the power-normalized optical force density and $\langle\mathbf{f},\mathbf{u}\rangle = \int \mathbf{f}^*\cdot\mathbf{u}\,dA$. Note that $k_{\mathrm{eff}}$ is the effective stiffness *per unit length*. In the case of radiation pressure forces $\mathbf{f}_{\mathrm{rp}}$ we have [103]

$$\mathbf{f}_{\mathrm{rp}} = \frac{1}{2}\left(\Delta\epsilon|\mathbf{e}_{\|}|^2 - \Delta\epsilon^{-1}|\mathbf{d}_{\perp}|^2\right)\hat{\mathbf{n}}\delta(\mathbf{r}-\mathbf{r}_{\mathrm{boundary}})$$

with $\delta(\mathbf{r}-\mathbf{r}_{\mathrm{boundary}})$ a spatial delta-distribution at the waveguide boundaries. The fields $\mathbf{e}$ and $\mathbf{d}$ are power-normalized. Here we already assumed that the Stokes and pump field profiles are nearly identical, which holds for intra-modal SBS given the small frequency shifts. Hence, we get

$$\langle\mathbf{f}_{\mathrm{rp}},\mathbf{u}\rangle = \frac{1}{2}\oint dl\,(\mathbf{u}\cdot\hat{\mathbf{n}})\left(\Delta\epsilon|\mathbf{e}_{\|}|^2 - \Delta\epsilon^{-1}|\mathbf{d}_{\perp}|^2\right) \tag{D4}$$

Additionally, the guided optical power $P$ is given by

$$P = \frac{v_{\mathrm{g}}}{2}\langle\mathbf{E},\epsilon\mathbf{E}\rangle = \frac{v_{\mathrm{g}}}{2}\int dA\epsilon|\mathbf{E}|^2 \tag{D5}$$

Combining equations (D2), (D4) and (D5), we find

$$\frac{\partial\omega_{\mathrm{c}}}{\partial x} = \frac{v_{\mathrm{g}}\omega_{\mathrm{p}}}{2}\langle\mathbf{f}_{\mathrm{rp}},\mathbf{u}\rangle$$

A similar derivation can be done for the strained bulk, so we have

$$\begin{aligned}\frac{\partial\omega_{\mathrm{c}}}{\partial x} &= \frac{v_{\mathrm{g}}\omega_{\mathrm{p}}}{2}\langle\mathbf{f},\mathbf{u}\rangle \\ \implies \langle\mathbf{f},\mathbf{u}\rangle &= \frac{2}{v_{\mathrm{g}}\omega_{\mathrm{p}}}\frac{\partial\omega_{\mathrm{c}}}{\partial x}\end{aligned} \tag{D6}$$

with $\mathbf{f} = \mathbf{f}_{\rm rp} + \mathbf{f}_{\rm es}$ and $\mathbf{f}_{\rm es}$ the electrostrictive force density. Substituting equation (D6) in (D3) yields

$$\tilde{\mathcal{G}} = \frac{2Q_{\rm m}}{\omega_{\rm p} v_{\rm g}^2 k_{\rm eff}} \left(\frac{\partial \omega_{\rm c}}{\partial x}\right)^2 \tag{D7}$$

Finally, we use the definition of the zero-point coupling rate $g_0 = x_{\rm ZPF} \frac{\partial \omega_{\rm c}}{\partial x}$ and the zero-point motion $x_{\rm ZPF} = \sqrt{\frac{\hbar}{2m_{\rm eff} L \Omega_{\rm m}}}$ with $m_{\rm eff}$ the effective mass *per unit length*. Inserting these in (D7) yields

$$\begin{aligned} \tilde{\mathcal{G}} &= \frac{2Q_{\rm m}}{\omega_{\rm p} v_{\rm g}^2 k_{\rm eff}} \frac{2m_{\rm eff} L \Omega_{\rm m}}{\hbar} g_0^2 \\ &= Q_{\rm m} \frac{4L}{(\hbar\omega_{\rm p})\,\Omega_{\rm m}} \frac{g_0^2}{v_{\rm g}^2} \end{aligned} \tag{D8}$$

and (D8) is identical to (D1). In this derivation, we started from full-vectorial definitions that are only valid for intra-modal forward scattering. In contrast, the mean-field transition shows that this result remains true with $v_{\rm g} \to \sqrt{v_{\rm p} v_{\rm s}}$ for inter-modal coupling.

#### *e. From independent derivative definitions*

The cavity resonance condition is $k_{\rm p} L = 2\pi m$ with $m$ an integer. Given $k_{\rm p} = \frac{\omega_{\rm p} n_{\rm eff}}{c}$ and $c$ the speed of light, this implies that

$$\left.\frac{\partial \omega_{\rm p}}{\partial x}\right|_{k_{\rm p}} = -\frac{\omega_{\rm p}}{n_{\rm eff}} \left.\frac{\partial n_{\rm eff}}{\partial x}\right|_{k_{\rm p}}$$

This can be recast in terms of the index sensitivity at fixed frequency by

$$\left.\frac{\partial n_{\rm eff}}{\partial x}\right|_{k_{\rm p}} = \frac{n_{\rm eff}}{n_{\rm g}} \left.\frac{\partial n_{\rm eff}}{\partial x}\right|_{\omega_{\rm p}}$$

with $v_{\rm ph} = \frac{c}{n_{\rm eff}}$ the phase velocity and $n_{\rm g} = \frac{c}{v_{\rm g}}$ the group index. Thus we have

$$\left.\frac{\partial \omega_{\rm p}}{\partial x}\right|_{k_{\rm p}} = -\frac{\omega_{\rm p}}{n_{\rm g}} \left.\frac{\partial n_{\rm eff}}{\partial x}\right|_{\omega_{\rm p}} \tag{D9}$$

The cavity frequency pull must be calculated at fixed wavevector ($g_0 = x_{\rm ZPF} \left.\frac{\partial \omega_{\rm p}}{\partial x}\right|_{k_{\rm p}}$), so this yields

$$\left(\left.\frac{\partial n_{\rm eff}}{\partial x}\right|_{\omega_{\rm p}}\right)^2 = g_0^2 \left(x_{\rm ZPF} \frac{\omega_{\rm p}}{n_{\rm g}}\right)^{-2} \tag{D10}$$

Previously [23], we showed that

$$\tilde{\mathcal{G}} = 2\omega_{\rm p} \frac{Q_{\rm m}}{k_{\rm eff}} \left(\frac{1}{c} \left.\frac{\partial n_{\rm eff}}{\partial x}\right|_{\omega_{\rm p}}\right)^2 \tag{D11}$$

Substitution of (D10) in (D11) with $x_{\rm ZPF} = \sqrt{\frac{\hbar}{2m_{\rm eff} L \Omega_{\rm m}}}$ results in

$$\tilde{\mathcal{G}} = \frac{4L Q_{\rm m}}{\hbar \omega_{\rm p} v_{\rm g}^2 \Omega_{\rm m}} g_0^2$$

or the other way around

$$g_0^2 = v_{\rm g}^2 \frac{(\hbar\omega_{\rm p})\,\Omega_{\rm m}}{4L} \left(\frac{\tilde{\mathcal{G}}}{Q_{\rm m}}\right) \tag{D12}$$

This proof only holds for forward intra-modal scattering – whereas the mean-field transition applies to backward and inter-modal scattering as well.

### Appendix E: Full solution of guided-wave evolution

In this section, we give the full solution of the traveling-wave spatial dynamics (8) in the constant pump approximation ($\Phi_{\rm p}(z) = \Phi_{\rm p}(0)$) where the pump is strong enough to be treated classically. From this solution, one can derive the regimes treated in section IV as limiting cases. In addition, this solution can directly be mapped on the cavity-based temporal dynamics (26). Thus, we start from

$$\begin{aligned} \partial_z a_{\rm s} &= \mp i \tilde{g}_0 b^\dagger a_{\rm p} \mp \tilde{\chi}_{\rm s}^{-1} a_{\rm s} \\ \partial_z b &= -i \tilde{g}_0 a_{\rm s}^\dagger a_{\rm p} - \tilde{\chi}_{\rm m}^{-1} b \end{aligned} \tag{E1}$$

which immediately yields

$$\left(\partial_z + \tilde{\chi}_{\rm m}^{-\star}\right)\left(\partial_z \pm \tilde{\chi}_{\rm s}^{-1}\right) a_{\rm s}(z) = \pm \tilde{g}^2 a_{\rm s}(z) \tag{E2}$$

where $\pm$ stands for forward (+) and backward (−) scattering. Inserting the ansatz $a_{\rm s}(z) \propto e^{\gamma z}$ leads to

$$\gamma^2 + \left(\tilde{\chi}_{\rm m}^{-\star} \pm \tilde{\chi}_{\rm s}^{-1}\right)\gamma \pm \left(\tilde{\chi}_{\rm m}^{-\star}\tilde{\chi}_{\rm s}^{-1} - \tilde{g}^2\right) = 0 \tag{E3}$$

Its solution is

$$\begin{aligned} \gamma_1 &= \frac{1}{2}\left\{-\left(\tilde{\chi}_{\rm m}^{-\star} \pm \tilde{\chi}_{\rm s}^{-1}\right) + \sqrt{\left(\tilde{\chi}_{\rm m}^{-\star} \mp \tilde{\chi}_{\rm s}^{-1}\right)^2 + 4\tilde{g}^2}\right\} \\ \gamma_2 &= \frac{1}{2}\left\{-\left(\tilde{\chi}_{\rm m}^{-\star} \pm \tilde{\chi}_{\rm s}^{-1}\right) - \sqrt{\left(\tilde{\chi}_{\rm m}^{-\star} \mp \tilde{\chi}_{\rm s}^{-1}\right)^2 + 4\tilde{g}^2}\right\} \end{aligned} \tag{E4}$$

Given these two roots, one can determine the exact evolution of the photon-phonon fields along the waveguide if the correct boundary conditions are known. The boundary condition $b(0) = 0$ and fixed probe flux $\Phi_{\rm s}(0) = a_{\rm s}^\dagger(0) a_{\rm s}(0)$ are appropriate for forward scattering such that

$$\begin{aligned} a_{\rm s}(z) &= \frac{a_{\rm s}(0)}{\gamma_2 - \gamma_1}\left\{\left(\gamma_2 + \tilde{\chi}_{\rm s}^{-1}\right)e^{\gamma_1 z} - \left(\gamma_1 + \tilde{\chi}_{\rm s}^{-1}\right)e^{\gamma_2 z}\right\} \\ b^\dagger(z) &= i\frac{a_{\rm s}(0)}{\tilde{g}} \frac{\left(\gamma_2 + \tilde{\chi}_{\rm s}^{-1}\right)\left(\gamma_1 + \tilde{\chi}_{\rm s}^{-1}\right)}{\gamma_2 - \gamma_1}\left\{e^{\gamma_1 z} - e^{\gamma_2 z}\right\} \end{aligned} \tag{E5}$$

The backward case (fixed $\Phi_s(L)$ with $L$ the waveguide length) can be solved similarly. This full solution contains the important regimes discussed in section IV. For instance, in the strong coupling regime ($\tilde{g} \gg \alpha_s + \alpha_m$) and at resonance ($\tilde{\Delta}_j = 0$) one can show that

$$\gamma_1 \approx -\frac{\alpha_m + \alpha_s}{4} + \tilde{g} \xrightarrow{\text{large } \tilde{g}} \tilde{g} \qquad \text{(E6)}$$
$$\gamma_2 \approx -\frac{\alpha_m + \alpha_s}{4} - \tilde{g} \longrightarrow -\tilde{g}$$

Therefore, the spatial coupling rate $\tilde{g}$ must overcome the average photonic and phononic propagation loss before actual photon-phonon pair generation can be seen. The photons and phonons indeed equally share the total propagation loss $\alpha_m + \alpha_s$ in this regime, as in cavity settings [12]. The spatial evolution (E5) then becomes identical to (30). The weak coupling regimes of stimulated photon ($\alpha_s \ll \alpha_m$) and phonon ($\alpha_m \ll \alpha_s$) emission can equally well be obtained from the full solution (E5). This solution also contains acoustic ($\alpha_s \ll \alpha_m$) [65] and optical ($\alpha_m \ll \alpha_s$) build-up effects.